\documentclass[11pt]{article}%
\usepackage[utf8]{inputenc}

\usepackage{natbib}
\setcitestyle{maxnames=1}
\usepackage{amssymb}
\usepackage{amsfonts}
\usepackage{amsmath}
\usepackage{comment}
\usepackage[nohead]{geometry}
\usepackage{setspace}
\usepackage[bottom]{footmisc}
\usepackage{graphicx}
\usepackage{float}
\usepackage{rotating}
\usepackage{gensymb}
\usepackage{lscape}
\usepackage{caption}
\usepackage{subcaption}
\usepackage{siunitx}
\usepackage{todonotes}
\usepackage{tabularx,booktabs}
\usepackage[small,compact]{titlesec}
\usepackage[hidelinks]{hyperref}
\usepackage{threeparttable}
\usepackage{color}
\usepackage{booktabs}
\usepackage[affil-it]{authblk}

\title{My Advisor, Her AI and Me: Evidence from a Field Experiment on Human-AI Collaboration and Investment Decisions\\[1ex]
\large Forthcoming in \textit{Management Science}}

\author[1]{Cathy (Liu) Yang}
\author[2]{Kevin Bauer}
\author[1]{Xitong Li}
\author[2]{Oliver Hinz}

\affil[1]{Department of Information Systems and Operations Management, HEC Paris}
\affil[2]{Department of Business Informatics and Information Economics, Goethe University Frankfurt}

\date{\today}

\begin{document}

\maketitle

\bigskip

\begin{abstract}
Amid ongoing policy and managerial debates on keeping humans in the loop of AI decision-making processes, we investigate whether human involvement in AI-based service production benefits downstream consumers. Partnering with a large savings bank in Europe, we produced pure AI and human-AI collaborative investment advice, which we passed to the bank customers and investigated the degree of their advice-taking in a field experiment. On the production side, contrary to concerns that humans might ineﬀiciently override AI output, our findings show that having a human banker in the loop of AI-based financial advisory by giving her the final say over the advice provided does not compromise the quality of the advice. More importantly, on the consumption side, we find that the bank customers are more likely to align their final investment decisions with advice from the human-AI collaboration, compared to pure AI, especially when facing more risky investments. In our setting, this increased reliance on human-AI collaborative advice leads to higher material welfare for consumers. Additional analyses from the field experiment along with an online controlled experiment indicate that the persuasive efficacy of human-AI collaborative advice cannot be attributed to consumers' belief in increased advice quality resulting from complementarities between human and AI capabilities. Instead, the consumption-side benefits of human involvement in the AI-based service largely stem from human involvement serving as a peripheral cue that enhances the affective appeal of the advice. Our findings indicate that regulations and guidelines should adopt a consumer-centric approach by fostering environments where human capabilities and AI systems can synergize effectively to benefit consumers while safeguarding consumer welfare. These nuanced insights are crucial for managers who face decisions about offering pure AI versus human-AI collaborative services and also for regulators advocating for having humans in the loop.

\end{abstract}

\newpage
\doublespacing

\section{Introduction}
Rapid technological advances in Artificial Intelligence (AI) have enabled machines to outperform human experts across an increasing array of tasks, including sales forecasting \citep{rohaan2022using}, stock mispricing detection \citep{avramov2022machine}, medical diagnosis \citep{goh2024large}, and the personalization of marketing interventions \citep{von2024smart}. Although industries continue to offer human-based services, businesses increasingly look at pure AI-based services as a replacement for human labor \citep[e.g.,][]{brynjolfsson2018can} due to enhanced productivity with reduced operational costs \citep[e.g.,][]{acemoglu2019automation}. 

However, pure AI services are not without their perils due to their inability to handle infrequent changes in input \citep{bauer2024feedback}, keep pace with shifts in latent data generation processes, and ensure decision accountability \citep{bergstein2020ai, cao2021man, feuerriegel2022bringing}. Consequently, policymakers, practitioners, and researchers frequently advocate for keeping humans in the loop of AI service provision---a proposition we refer to as ``human-AI collaboration" in this paper. %\citep{buckley2021regulating,cowgill2020algorithmic}. 
For instance, Article 22 of the EU's General Data Protection Regulation (GDPR) promotes ``human intervention" in delivering AI-based services, thereby regulating decision-making processes that rely solely on automated algorithms. More recently, Article 14 of the EU's AI Act requires that humans must be capable of overseeing high-risk AI systems to mitigate risks to people's health, safety, or fundamental rights. Relatedly, Vanguard Personal Advisor, an investment advisory service, differentiates itself from its competitor, Wealthfront, by offering clients AI-based financial advice complemented by access to human experts rather than providing fully automated services.%%%
%%%
%%%
\footnote{See \url{https://investor.vanguard.com/advice/personal-hybrid-robo-advisor} and \url{https://www.wealthfront.com/robo-advisor-investing}}
%%%
%%%

Despite the considerable practical and regulatory implications, only a few recent studies have systematically examined the ramifications of keeping humans in the loop and its impact on AI-based service quality \citep[e.g.,][]{fugener2021will}. However, businesses and policymakers must assess the costs and benefits of integrating humans into the AI-based service value creation process, encompassing both production and consumption aspects \citep{bauer2024all,hermosilla2018can}. To the best of our knowledge, no prior research examines the effect of human involvement in AI-based service delivery on downstream consumption. Thus, the present paper aims to fill this literature gap by examining AI-based service production with (versus without) human involvement and subsequently studying its downstream consumption.

From a theoretical standpoint, the impact of human involvement in AI-based services on its consumption remains ambiguous. On the one hand, consumers might interpret the presence of humans as enhancing service quality, either through synergistic human-AI interactions \citep{lu20241+} or by receiving emotional reassurance from direct human involvement in the service process \citep{loewenstein2001risk}. On the other hand, human intervention could be perceived as introducing noise that diminishes service quality compared to fully automated AI solutions \citep{kahneman2016noise}, or it might trigger negative reactions if viewed as a persuasive tactic \citep{friestad1994persuasion}. Consequently, a systematic investigation of these competing predictions %and the underlying mechanisms 
is essential for firms and policymakers to assess the relative persuasive effectiveness of the human-AI collaborative compared to pure AI solutions. Additionally, understanding why human involvement in AI-based services would influence service consumption is similarly critical as it can inform managerial and regulatory decisions of what human involvement needs to look like and how it should be communicated. 

This paper examines the consumption-side effects of keeping humans in the loop, particularly when businesses---today or in the future---can choose between fully automated AI-driven services and human-AI collaborative solutions. We pose the following main research questions:\vspace{0.2cm}

\textit{\textbf{RQ1}: Does human involvement in the AI-based service provision influence downstream consumers' response to the service ultimately produced?
}

\textit{\textbf{RQ2}: What is the primary underlying mechanism that drives the possible effect?
}

\vspace{0.2cm} To address our research questions, we partnered with one of Germany's largest savings banks to investigate the consequences of human-AI collaborative financial advice. The financial services context is particularly well suited for this investigation for at least two reasons. First, many services in this industry, subject to regulatory constraints, are already amenable to automation by AI, making the choice between fully automated AI-driven services and human-AI collaborative offerings especially pertinent. In the context of financial advisory service, the focus of our empirical study, a survey by Vanguard, suggests that AI-based advisory already made up 23\% of the market in 2022 \citep{costa2022quantifying}, with an expectation to grow at an approximate compound annual rate of 30\% \citep{marketsandmarketsFinanceMarket}.%%%
\footnote{For example, providers such as \textit{\href{https://www.portfoliopilot.com/}{PortfolioPilot}} and \textit{\href{https://www.sofi.com/}{SoFi Invest}} employ AI to deliver fully automated financial advice. Comparable automation is evident in areas such as loan approval \citep{faisal2021credit}, insurance claims processing \citep{singh2019automating}, and retirement planning \citep{wasserbacher2022machine}.}
%%%
%%%
Consequently, financial service providers are, or will soon be, particularly likely to confront the decision between implementing pure AI solutions and adopting human-AI collaborative models. Second, financial decisions significantly affect personal wealth and broader economic outcomes, making AI deployment in the financial services sector a relevant decision from a societal and regulatory perspective. In this context, understanding the consumption-side effects of human oversight in AI-based services is essential for determining whether the intended protective measures promote overall welfare or inadvertently introduce inefficiencies related to specific consumer responses.

In our research context, the bank explored the introduction of a new investment vehicle---personal loan investments---to its customers and provided them with investment advice to assist their investment decisions. This financial advisory service could already be offered fully AI-based from a technical perspective \citep[e.g.,][]{chang2022machine}. However, this is not necessarily the default as of yet, providing a unique opportunity to study consumption-side effects of keeping humans in the loop. Together with our industry partner, we designed and conducted a two-staged study in which we had control over the investment advice production process in Stage 1, enabling us to observe its consumption in Stage 2 via an experimental design. Specifically, in Stage 1 of the study, we developed a machine learning model that generates investment advice as a pure AI service. We then presented this AI-based advice to bank advisors (hereafter referred to as ``bankers'') to produce the human-AI collaborative investment advice. The results of this stage show that humans in the loop significantly change the type of advice but do not compromise its average quality. In Stage 2 of the study, we passed the real AI and human-AI collaborative investment advice from Stage 1 to the bank's customers and gauged their responses in a randomized and incentivized field experiment. To gain additional insights into the underlying mechanisms, we created an additional experimental condition by providing customers the human-AI collaborative advice without revealing the bankers' use of AI and labeled it as ``human advice." This additional control treatment allows us to tease out whether the potential treatment effect of the human-AI collaborative (vs. AI) advice stems from consumers' inferences about any change in the service quality via the interaction of two agents (banker and AI) or the mere involvement of human bankers in the human-AI collaborative advice production process.%%%
%%%
%%%
\footnote{We could also create an experimental condition where the human-AI collaborative advice is provided without revealing the bankers' participation. However, because bankers did have the final say in producing the human-AI collaborative advice, concealing such facts may run into ethical risks. Thus, it is not practically feasible in the field setting.} 
%%%
%%%

In Stage 2 of the field study, the core of this paper, we approached bank customers during their visits to the branch, inviting them to invest in the new investment vehicle of personal loans with financial endowments. After each customer made initial investment decisions and risk assessments for available investment opportunities, we randomly assigned each customer to receive investment advice from either an AI (baseline), a human-AI collaboration (main treatment), or a human-AI collaboration labeled as human (additional control), to make final decisions. Our experimental results show that bank customers exhibit greater alignment in their final investment decisions with advice from human-AI collaboration than from AI. Interestingly, there is no significant difference in the alignment of investment with advice from a human-AI collaboration and a human alone, suggesting that the customers' greater reliance on human-AI collaborative advice is unlikely driven by their inferences about greater service quality resulting from the interaction of two agents (banker and AI). 

Further subsample analyses suggest that the increased reliance on advice is more pronounced for relatively risky investments which entails greater uncertainty. We interpret this heterogeneity through the lens of the classical elaboration likelihood model (ELM) from the persuasion literature. This theoretical framework posits that persuasion could operate via the central and peripheral routes \citep{petty1986elaboration,tversky1974judgment}. The central route involves carefully considering the content and quality of the information presented. In contrast, the peripheral route relies on non-informative cues such as the emotional appeal of the source. While the two persuasion routes could operate simultaneously, the peripheral route typically dominates the central route with growing decision uncertainty \citep{petty1986elaboration}. 

To further uncover the underlying mechanisms, we conducted an additional controlled and incentivized online experiment. We do not find evidence that the greater persuasive effectiveness of human-AI advice, compared to AI advice, is driven by the central route. Instead, the results from the online experiment provide process evidence supporting that the bank customers' positive response to a banker's involvement in AI-based solutions is primarily driven by increased emotional trust in the advice source via the peripheral route, especially when experiencing heightened investment uncertainty.

Our paper makes several key contributions to the literature. First, we advance research on human-in-the-loop systems \citep{fugener2021cognitive, lebovitz2022engage} by empirically examining the downstream effects of human-AI collaboration in financial services. Using evidence from a field experiment with a realistic service provision scenario, we demonstrate how incorporating human involvement in AI-driven financial advisory influences its consumption and increases downstream customers' welfare. Second, we address the literature gap by showing consumers' positive response to human involvement in AI-based services, which is primarily derived from social cues that engender affective decision comfort. Third, consumers' greater reliance on human involvement in the AI-based advice service suggests a potential remedy for algorithm aversion when AI-based advice is beneficial \citep[e.g.,][]{longoni2019resistance}. In particular, allowing humans to have the final say in the AI-based advisory service production process can mitigate downstream consumers' aversion to algorithmic advice---for better or worse.

Our findings have important implications for practitioners and policymakers, highlighting that the influence of humans in the loop of AI-based services goes beyond affecting the type of advice produced upstream in the value chain. It also impacts the downstream consumption of the generated advice, especially when consumers experience high uncertainty. Our findings highlight that the role of humans in AI-driven service production is not merely supplementary but may be a critical determinant of whether AI-enabled eﬀiciencies translate into tangible welfare gains for end consumers. This suggests a need for a more holistic approach to AI deployment, where the potential of human-AI collaboration enhancing service quality and consumer trust can be fully realized. For policymakers, our findings indicate that regulations and guidelines should also focus on fostering environments where human capabilities and AI systems can synergize effectively to benefit consumers while safeguarding consumer welfare. 

\section{Related Literature and Theoretical Background}

\subsection{Human in the Loop of AI-Based Solutions \label{human-in-the-loop}}

AI outperforms humans in many prediction tasks \citep{agrawal2022prediction,cowgill2020algorithmic,rahwan2019machine}. However, fully automated AI-based solutions may not always be desirable due to AI's inherent limitations, such as its inability to reason \citep{bergstein2020ai}, adapt to infrequent input changes \citep{cao2021man}, and provide decision accountability \citep{feuerriegel2022bringing}. Consequently, there is a growing emphasis among policymakers and practitioners on human intervention in the production of AI-based services \citep{kellogg2020algorithms}. This approach aims to enhance performance, fairness, and accountability \citep{buckley2021regulating,cowgill2020algorithmic}, while also addressing ethical and legal challenges associated with purely AI-driven services \citep{awad2018moral,kingston2016artificial}. Reflecting this perspective, the European Parliament's AI Act came into effect in August 2024 (with the first provisions becoming mandatory in February 2025), seeking to regulate AI-based products and services by defining the necessary degree of human intervention based on the potential risks to consumer welfare. Despite these regulatory and practical demands, our understanding of the implications of keeping humans in the loop of AI-based service offerings remains limited.

On the service production side, previous studies often find that human involvement can diminish the quality of AI-driven services. Specialized professionals, such as medical doctors \citep{lebovitz2022engage}, judges \citep{berk2017impact}, and supply chain workers \citep{sun2022predicting}, frequently discount AI-generated advice, possibly resulting in lower service quality compared to that produced by AI alone. Experts' tendency to discount AI advice aligns with the broader concept of ``algorithmic aversion," where humans deviate from algorithmic recommendations due to lower tolerance for algorithmic errors \citep{dietvorst2015algorithm}, insufficient trust in AI for subjective tasks \citep{castelo2019task}, the high-stakes nature of decision contexts such as financial investments \citep{onkal2009relative} and medical diagnoses \citep{longoni2019resistance}, or a perceived threat to their decision-making authority \citep{lebovitz2022engage}. Conversely, there are instances where over-reliance on AI can also degrade service outcomes. For example, \cite{dell2021falling} find that recruiters produce poorer resume screening results when they ``fall asleep at the wheel" by overly trusting AI's high performance. Similarly, \cite{fugener2021cognitive} report that humans may start to overly rely on AI so that existing complementarities in human and AI skills may disappear. \cite{bauer2023expl} observe that humans may overly adjust their decision to observed AI predictions, leading to lower task performance, especially when explanations accompany predictions. However, human intervention in AI-based solutions is not always detrimental. Recent research by \cite{lu20241+} suggests that humans can improve AI service quality when the information environment is well-calibrated and supported by clear AI explanations. Additionally, \cite{abdel2023ai} demonstrate that humans can benefit due to deeper reflections about their own way of reasoning and can learn from AI when they collaborate closely within the decision-making process.

Although existing literature suggests mixed evidence on how human involvement affects AI service quality on the service production side, businesses and policymakers must also evaluate the costs and benefits of incorporating humans into AI service consumption. Despite its importance, there is limited understanding of how downstream consumers would respond to human-AI collaborative services, compared to purely AI-driven ones. Recently, an emerging body of research investigates consumer responses to services co-produced by humans and generative AI (genAI), with and without disclosing genAI's involvement. These recent studies, which compare human-AI collaboration to human-only services, generally find that consumers tend to devalue products or services when they know that genAI played a role in their creation \citep{bauer2024all,darda2023computer,millet2023defending}. This discounting may stem from consumers' appreciation of human effort in service production. However, extrapolating these findings to understand consumer reactions to human-AI collaborative services compared to purely AI-driven services presents challenges. First, while studies contrasting human-AI collaboration with human-only services focus on consumer perceptions of AI's contribution to the collaboration, our research shifts the focus to understanding consumers' reactions to the human contribution within a human-AI collaborative service by comparing it to an AI-only service. Humans offer social attributes that AI lacks, such as shared responsibility \citep{feuerriegel2022bringing} and emotional reassurance during moments of decision discomfort \citep{komiak2006effects,loewenstein2001risk,mcknight2002developing}. These unique human qualities suggest that consumer reactions to human involvement in AI-based services may differ significantly from their reactions to AI involvement. Second, the current findings on consumer discounting of human-AI collaboration relative to human-only services do not clarify whether this discount arises from the mere presence of AI or a nuanced appreciation of human involvement relative to AI. The former does not provide insight into consumer responses to human participation in AI-based services, while the latter implies a potential positive response to human involvement when compared to pure AI. Additionally, it is challenging to apply these findings from creative tasks, which involve subjective evaluations \citep{bauer2024all,millet2023defending}, to financial investment contexts, where decisions are driven by objective outcomes. Given previous research suggesting that task objectivity enhances the value placed on AI over human services \citep{castelo2019task}, one might anticipate that consumers could react negatively to human intervention in AI-based financial advisory services.

Given the lack of evidence on the downstream ramifications of keeping humans in the loop of AI advice production, we focus on understanding whether consumers react positively or negatively to this human involvement compared to pure AI-based service and examine the underlying mechanism for this preference.

\subsection{Persuasive Efficacy via the Central and Peripheral Routes}\label{persuasion}

To explore the role of human involvement in AI-based advice, we employ the ELM from the persuasion literature \citep{petty1986elaboration}. This framework posits that having a human in the advice production process can act as a persuasive element, influencing consumers' willingness to incorporate the advice and balance against their own judgment, which can eventually lead to a change in decision outcomes. This change of judgment and decision outcomes is considered persuasive efficacy, frequently measured by the degree of advice-taking \citep{bonaccio2006advice}. 

According to the ELM, persuasion operates through two primary, non-exclusive routes: the central route and the peripheral route. The central route of persuasion involves a high level of cognitive engagement, where individuals carefully evaluate the substance of the information presented, such as its completeness, accuracy, and logical coherence. The peripheral route typically operates on a lower cognitive level and is driven by non-content-related cues, such as the emotional appeal of the advice source. While the two persuasion routes could operate simultaneously, the peripheral route normally interrupts the central route of information processing with growing decision complexity \citep{petty1986elaboration} and experienced decision discomfort \citep{loewenstein2001risk}.

In the context of financial decision-making, the process of making an investment demands individuals to make well-informed choices, given the monetary implications involved and the need to consider a substantial amount of information. As a result, individuals may predominantly rely on the central persuasion route to form an investment decision by deliberately evaluating its related information. However, all financial investments entail uncertainty regarding their potential returns, leading to increased perceived decision complexity. This may, in turn, prompt individuals to engage in the peripheral route of information processing when making an investment decision \citep{tversky1974judgment}. 

We aim to understand to what extent the persuasive effect of human involvement in AI-based advice occurs through the central and the peripheral routes. This differentiation in the mechanism has implications for how banks and other organizations should implement human-in-the-loop communication strategies. If persuasion primarily occurs through the central route, it would be crucial to communicate the genuine predictive performance of human-AI collaborative advice, helping consumers form realistic expectations about the advice they receive. Conversely, if the peripheral route dominates, careful management of the cues introduced by human advisors becomes essential to avoid unintended influences on consumers' decisions. In the following, we discuss in detail how human involvement in AI-based service could change the persuasive efficacy via the central and peripheral routes and link these arguments to existing literature.

\subsubsection{Central Persuasion Route}

In the financial investment context, persuasion through the central route occurs when human involvement in the AI-based advice production process prompts consumers to reassess the quality of the advice. The impact of human involvement through this route can vary; it may either enhance or diminish consumers' adherence to the advice, depending on their perceptions of the human's contribution to the overall quality of the advice \citep{petty1986elaboration}. The literature offers evidence supporting both potential outcomes.

On the one hand, human involvement could enhance persuasion through the central route by increasing cognitive trust in the overall advice-giving system. Consumers may expect that human judgment, particularly when combined with AI capabilities, leads to more trustworthy advice regarding its coherence, overall competence, and benevolence \citep{komiak2006effects}. Relatedly, consumers may expect human-AI collaborative advice to be more accurate than AI due to complementarities \citep{lu20241+}. Such changes in beliefs may be grounded in the notion that humans excel in areas where AI might fall short, such as understanding context, applying ethical considerations, and exercising discretion in ambiguous situations \citep{castelo2019task}. 

On the other hand, there is the possibility that human involvement could trigger skepticism, thereby reducing the effectiveness of the central route to persuasion. If consumers perceive that AI inherently outperforms humans in objective, data-driven tasks such as investment advisory, they may view human intervention as a detractor rather than an enhancer of advice quality \citep{castelo2019task}. This skepticism could be fueled by concerns that human judgment might introduce bias or noise into what would otherwise be a purely rational, data-driven process, thereby undermining the precision of the AI's recommendations \citep{kahneman2016noise}. In such cases, consumers might be less inclined to engage deeply with the advice, perceiving human involvement as a negative signal.

\subsubsection{Peripheral Persuasion Route}

In the context of AI-generated investment advice, consumers could also perceive peripheral cues, such as social and emotional components introduced by the presence of a human advisor. One critical aspect of the peripheral route is the role of human advisors in providing decision comfort. Decision-making often involves significant emotional elements, where feelings of anxiety and discomfort can impede effective choices, particularly in high-stakes or complex situations \citep{loewenstein2001risk}. The presence of a human advisor may alleviate these negative emotions through several mechanisms associated with the peripheral route of persuasion.

First, humans in the loop may offer a sense of decision accountability, making consumers feel more at ease knowing that a human, rather than an AI, shares the responsibility for the outcome of the decision. This shared accountability can alleviate the burden of making difficult choices, as the human advisor can be seen as a figure who will bear some of the blame if the decision leads to a negative outcome \citep{feuerriegel2022bringing}. The mere presence of this accountability cue can reduce consumer anxiety and increase their willingness to follow the advice. Second, humans in the loop may foster emotional trust, a critical factor in persuasion via the peripheral route. Unlike AI, human advisors can provide empathy, understanding, and reassurance—qualities that help build a relational connection with consumers. Emotional trust may enhance the appeal of the advice, making consumers more likely to accept and act on it \citep{komiak2006effects}. 
Third, the presence of a human advisor may introduce social pressure, subtly encouraging consumers to conform to the advice provided. This social influence can lead consumers to relinquish some degree of decision autonomy, as they may feel socially compelled to align their choices with the advice they know to, at least partially, originate from a human source \citep{asch1955opinions}.

Given that the abovementioned three factors are well-documented contributors to an advisor's persuasive efficacy \citep{castelo2019task,harvey1997taking,koestner1999follow,sniezek1995cueing}, it seems plausible to assume that the involvement of a human banker in the advice production process could lead to an increase in downstream customers' advice taking, as these social cues are likely to enhance the overall peripheral persuasiveness of the advice. However, it is also important to consider the potential downside: consumers may become wary of the human advisor's influence if they perceive the presence of a human as a deliberate persuasion attempt rather than a genuine effort to assist, which could undermine the effectiveness of the advice \citep{friestad1994persuasion}.

\section{Research Context and Study Overview}\label{research_context}

We partnered with one of the largest savings banks in Germany, a prominent financial institution operating thousands of branches and serving millions of customers, to conduct the study. In 2021, the bank began exploring the introduction of a new investment vehicle---personal lending---to its customer base. In our experiment, we utilized real personal loans from one of the largest peer-to-peer lending platforms as genuine investment opportunities for customers. Specifically, we accessed a historical dataset that included lending outcomes and seven key pieces of information provided by borrowers: loan amount, duration (in months), annual percentage rate (APR), monthly installments, borrowers' annual income, occupation, and the purpose of the loan.

Personal lending presents a compelling investment option due to its potential for higher returns compared to other fixed-income opportunities, while it also carries notable credit risks. Given this inherent risk, offering sound advice on the creditworthiness of personal loan applicants is critical for the bank to safeguard its customers' financial interests. In line with the bank's standards, the investment advice for each loan consisted of an investment recommendation for the typical customer (i.e., whether it is profitable to invest or not) and a default risk assessment categorized into seven levels, ranging from extremely low to extremely high risk. These risk levels align with the bank's traditional classification system for assessing default risks.

Our partner and we were particularly interested in leveraging AI to automate investment advisory for personal lending, assuming that AI typically offers higher prediction accuracy and consistency at a lower cost than human advisors \citep{kahneman2016noise}. This setting provided a unique opportunity to explore how keeping a human in the loop and giving her the final say in an AI-driven advice production might influence both the advice production and consumers' advice consumption in a real-world business setting.

\begin{figure}
    \centering
    \caption{Field experimental design and procedure overview\label{fig:field_experimental_design}}
    \includegraphics[width=0.7\linewidth]{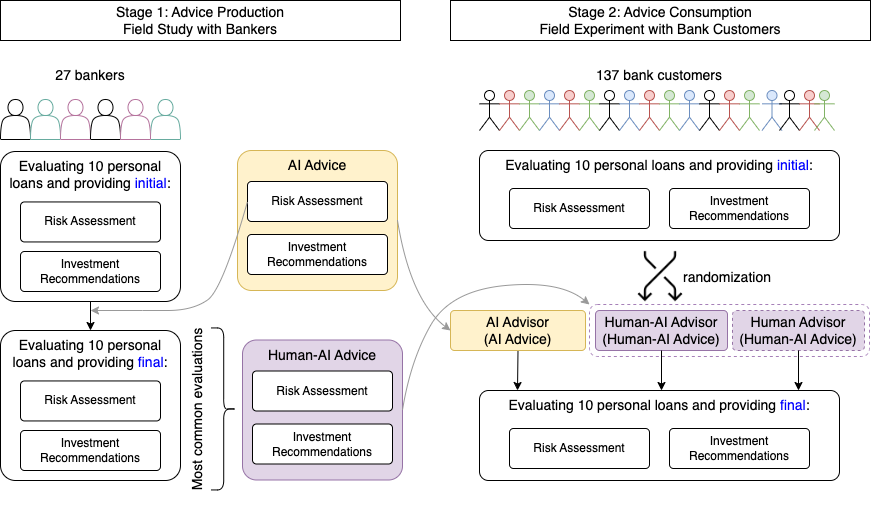}   
\end{figure}

We designed a two-staged study to test whether (i) the inclusion of a human banker in the AI-driven advice production process leads to any change in the advice ultimately produced and (ii) this human involvement affects downstream customers' advice consumption. The experiment was conducted across two geographically proximate bank branches with similar consumer demographics. 
Figure \ref{fig:field_experimental_design} provides an overview of the study design and the main procedures involved in both the production and consumption of human-AI collaborative and AI-only advice.
In Stage 1, we allowed bankers to interact with an AI system that we had previously trained to generate investment advice tailored to the typical consumer profiles in their branch. To assess if bankers did take AI input into account, we employed a well-established methodology \citep[e.g.,][]{bonaccio2006advice}, involving two rounds of advice production: initially without AI assistance, and subsequently in collaboration with the AI. This stage was crucial to understand whether financial experts would alter the AI-generated advice when collaborating with the AI-based system. In Stage 2, we utilized the investment advice produced in Stage 1 to conduct a randomized field experiment with real bank customers who were presented with various personal lending options available for investment. Using a between-subjects design, we exogenously varied the source of the investment advice, randomly assigned from pure AI, human-AI collaboration, or human advisor (collaborative advice with AI involvement being undisclosed).  
Analogue to Stage 1, we measured consumers' engagement with the advice through two consecutive rounds of decision-making. 

%%%%%%%%%%%%%%%%%%%%%%%%%%%%%%%%%%%%%%%%%%%%%%%%%%%%%%%%%%
%%%%%%%%%%%%%%%%%%%%%%%%%%%%%%%%%%%%%%%%%%%%%%%%%%%%%%%%%%
%%%%%%%%%%%%%%%%%%%%%%%%%%%%%%%%%%%%%%%%%%%%%%%%%%%%%%%%%%
%%%%%%%%%%%%%%%%%%%%%%%%%%%%%%%%%%%%%%%%%%%%%%%%%%%%%%%%%%

\section{Field Study on the Production Side}\label{advice_generation}

In this section, we provide details on the AI we developed and how we used it in the first stage of our study with bankers to produce the human-AI collaborative advice.

\subsection{AI advice}\label{ai_advice_generation}

We developed a machine learning model to predict personal loan defaults (yes or no) using the seven key attributes of loan applicants described earlier. We began by randomly selecting 90\% of the available data on matured personal loans listed on the lending platform from its inception through the fourth quarter of 2018 ($N$ = 1,061,279) as our training set while reserving the remaining 10\% as the test set. Following best practices in data science \citep{pfeuffer2023explanatory}, we employed a five-fold cross-validation approach to empirically determine the optimal model architecture and corresponding hyperparameters. This process involved iteratively testing various configurations and assessing their performance to identify the most effective model for our specific context. After thorough empirical optimization, we identified a neural network as the best-performing model for predicting loan defaults in our setting.

We validated the model's effectiveness on the test data and found that the AI achieved an accuracy of 0.73 in predicting borrower defaults, with a Receiver Operating Characteristic Area Under the Curve (ROC-AUC) score of 0.71. The achieved performance is comparable to that of contemporary machine learning models in the domain of personal loan default prediction, confirming the reliability and effectiveness of our model.%%%
%%%
%%%
\footnote{Using the same data set, \citet{kumar2016credit} obtained the best accuracy of 88.5\% by using a random forest algorithm with a considerably vast number of features. \citet{malekipirbazari2015risk} used a random forest and reached an accuracy of 78\% using 13 features, and 69.8\% using 4 features. Note that with a considerably larger set of features \cite{chang2022machine}, that was prohibitive in our study, achieve a higher performance.}
%%%
%%%

We utilized the model's predicted default probabilities to generate both investment recommendations and risk assessments. In collaboration with the bank, we established specific guidelines for the AI's decision-making process: (i) the AI recommends an investment if the expected payoff is higher than the alternative of not investing, and (ii) it categorizes the associated risk according to the bank's standard risk classes. These risk categories are defined as follows: extremely low risk (less than 10\%), very low risk (10\% to less than 20\%), low risk (20\% to less than 40\%), moderate risk (40\% to less than 60\%), high risk (60\% to less than 80\%), very high risk (80\% to less than 90\%), and extremely high risk (greater than or equal to 90\%).

\subsection{Human-AI Collaborative Advice}

\subsubsection{Production Process}
To generate human-AI collaborative advice, we engaged human bankers with our developed AI model. We selected 24 personal loans (balanced in risk level and the actual default outcome) from the test data not included in the AI training as candidate investment opportunities to generate human-AI collaborative advice. Twenty-seven experienced bankers from the two bank branches participated in generating human-AI collaborative advice over a two-week period in September 2021. Considering bankers' time constraints, we developed an online interface and presented each banker with ten randomly selected personal loans from the 24 candidate investment opportunities. For every investment opportunity, bankers observed the seven loan characteristics, which were identical to the features that the AI used to predict credit default. Each banker made investment recommendations and risk assessments (ranging in seven levels from extremely low to extremely high) on each personal loan once before and once after receiving the risk assessment from the AI. We only presented the AI's risk assessments to the bankers without its investment recommendation, which helped to ensure that the bankers critically evaluated the AI's input rather than passively accepting the binary investment recommendation. Figure \ref{fig:stage1_interface} in the Online Appendix depicts the interface that instructed the bankers to make risk assessments and investment recommendations with the AI's risk assessment. 

We provided financial incentives to ensure bankers produce high-quality advice to align their interests with their customers. Specifically, each banker was compensated according to the accuracy of two randomly selected investment recommendations: one from their ten initial investment recommendations and the other from their ten final investment recommendations. Each correct investment recommendation (i.e., advice to invest when the loan did not default and not invest when it defaulted) yielded a reward of ten Euros. On average, bankers completed their recommendations in approximately ten minutes and earned an hourly wage of approximately eighteen Euros. 

We generate the human-AI collaborative advice for the subsequent bank customers based on the bankers' final risk assessments and investment recommendations they generated in collaboration with AI. We next test whether this human-AI collaborative advice differs from the AI advice.

\subsubsection{Influence of Human Involvement and AI in Human-AI Collaborative Advice}

Does human involvement influence AI-based advice production? We answer this question by examining how bankers and AI each contributed to the final human-AI collaborative advice. If the bankers made no contributions, we would expect the human-AI collaborative advice to closely align with the AI's original recommendations. Conversely, if the AI did not influence the collaborative advice, we would not expect any increase in the alignment between the bankers' risk assessments and investment decisions after they actually saw the AI's risk assessment. To quantify these dynamics, we introduce two key metrics: $GapRiskAssess$ and $InvestAlign$, where $GapRiskAssess$ captures the absolute difference %gap 
between the bankers' and the AI's risk assessments both before and after receiving the AI's risk assessment; $InvestAlign$ measures the alignment between the bankers' and the AI's investment recommendations, coded as one for alignment and zero otherwise, before and after receiving AI's risk assessment.%%%
%%%
\footnote{Note that although bankers were only provided with risk assessments without investment recommendations from the AI as information to update their judgment, we also look at investment alignment since AI's risk assessment negatively correlates with the investment recommendations which was also calculated based on the predicted default chance from AI.} 
%%%
%%%

We first investigated whether bankers contributed to the human-AI collaborative advice. One-sample Student's t-tests suggest that $GapRiskAssess$ after receiving AI advice on average is significantly greater than zero ($M=0.93$, $SD=0.91$, $N=270$, $p<0.001$) and $InvestAlign$ after receiving AI advice is on average significantly less than one ($M=0.70$, $SD=0.46$, $N=270$, $p<0.001$). These results show that human-AI collaborative advice involves bankers' own judgment and thus does not perfectly mirror AI advice. We then explored whether AI weighs into the human-AI collaborative advice by increasing bankers' alignment in the risk assessment and investment recommendations to that of AI. To test this, we regressed $GapRiskAssess$ and $InvestAlign$ on an independent variable indicating whether the banker made the investment recommendation after receiving AI advice, controlling for banker-loan-investment order three-way fixed effects. The results, presented in Table \ref{tab:stage1_ai_impact} in the Online Appendix, show that bankers' risk assessments and investment recommendations became significantly more aligned with those of the AI after receiving its advice. The evidence suggests that both bankers and AI have contributed to the production of human-AI collaborative advice. That is, we observe production-side effects: in our setting with professional bankers, human involvement influences the upstream advice production process, resulting in advice that differs from what the AI (or human banker) would have produced independently.

Notably, although it is not our primary variable of interest due to its endogenous dependence on the AI's prediction performance and the specific set of 24 loans selected, we observed that the involvement of bankers in the advice production process does not compromise the accuracy of the investment recommendations in our specific setting. We define a recommendation as accurate if it advises investing when the loan does not default and not investing when it does default. The difference in the investment recommendation accuracy between the 27 sets of human-AI collaborative advice generated from each banker-AI pair and the AI-only advice is not statistically significant ($M = -0.03$, $p = 0.23$, one-sample Student's t-test). 

\subsubsection{Human-AI Advice Curation}
Considering the cognitive burden and time constraints in evaluating all 24 personal loans assessed in Stage 1, we randomly selected the same subset of ten loans for all customers as potential investment opportunities. For each of the ten loans, we also provided identical human-AI collaborative and AI advice for the customers assigned under the same experimental condition. We did so for two reasons. First, the bank we collaborated with insisted on delivering identical advice from a banker-AI pair to all customers receiving human-AI collaborative advice to avoid potential customer dissatisfaction if they discovered that different advice was offered to others within the same experimental condition. Second, from a technical perspective, rather than randomly drawing advice from different bankers' investment recommendations and risk assessments when collaborating with AI in Stage 1, we presented the bank customers with the most frequently occurring human-AI collaborative advice for each loan from Stage 1 to reduce the extensive variation and noise in recommendations. This strategy could prevent the selection of unrepresentative human-AI advice and ensure that customers receive the advice they are most likely to encounter if paired with one randomly selected banker working alongside the AI for generalizability. Under this consideration, we informed the bank customers who received human-AI collaborative advice in the field that the advice was from one banker who worked with AI. We emphasize that the bank customers who later received human-AI collaborative advice were not aware of the number of bankers involved in the advice production study, nor did they learn that the advice was the most frequent advice among the bankers collaborating with AI, which could positively bias them to follow such advice.

We present the details of the ten loans, along with the investment recommendations from the AI and human-AI collaboration, in Table \ref{tab:loan_advice_details}. Seven of the ten investment recommendations from both the AI and human-AI collaboration are consistent with the customers' financial interest, such that a loan did not default when the recommendation was to invest and defaulted when the recommendation was not to invest. We note that among ten loans, AI and human-AI advice differ on six regarding risk assessments and two concerning investment recommendations. This leaves us with four loans with identical human-AI and AI risk assessments and investment recommendations. We decided to share the actual AI and human-AI investment advice with the bank customers, so they receive realistic advice in the field experiment.
In the analyses, we account for the advice differences contingent on a loan across conditions.%%%
\footnote{We acknowledge variations in how we communicate the advice source (see Table \ref{tab:field_source_stimuli}) could potentially confound the treatment effect. These additional variations in the treatments primarily stem from practical considerations in field implementations. In particular, we provided the customers in the AI-only and Human-AI conditions with the description of the AI and its prediction accuracy to the bank customers following the best practice suggested by the European Commission for the EU AI Act \citep{EU2024} back in 2021. We did not provide customers in the Human-only condition with the prediction accuracy of the banker, as it is uncommon for bank customers to demand the prediction accuracy of a banker given the feedback from the bank we collaborated with. Instead, we ensured the customers that the banker used their experience and had financial interest if they made good investments. To mitigate concerns that the differences in communication systematically bias our results, we present consistent results from a laboratory experiment in Section \ref{lab} that minimize these variations across experimental conditions.} To further address potential endogeneity concerns arising from such differences, we show consistent results from a laboratory experiment in Section \ref{lab} that provided identical advice to participants across all conditions.

\section{Field Experiment on the Consumption Side}

In the second stage of our field study, we sought to understand how customers' awareness of human involvement in AI-driven investment advice influences their decision-making. To this end, we conducted a field experiment with customers who visited the bank's branches in person.

\subsection{Experimental Design}\label{field_exp_design}

Our field experiment aims to understand the persuasive efficacy of human-AI collaborative compared to AI advice for bank customers. Our baseline condition is when the bank customers receive advice from the AI (``AI-only" condition hereafter), and our main treatment condition is when the bank customers receive advice from the human-AI collaboration (``Human-AI" condition hereafter). We created one additional control treatment by offering bank customers the human-AI collaborative advice without revealing the banker's use of AI (``Human-only" condition hereafter). 

We focus on estimating the impact of Human-AI (vs. AI) on customers' reliance on advice throughout the paper as the main result. We use the Human-only condition to understand the underlying mechanism of how banker's participation leads to the Human-AI (vs. AI) treatment effect. In particular, we anticipate a reduced magnitude of a positive (negative) treatment effect from the Human-only relative to that from the Human-AI condition if the bank customers believe the superiority (inferiority, respectively) of the human-AI collaborative advice is due to the complementarity (the bias introduction from an additional party, respectively) of the two advisors upon interaction (central route). In contrast, we expect no difference in the persuasion efficacy under the Human-AI and Human-only conditions if the banker's involvement in the AI-based recommendation does not solely influence the cognitive perception of multiple parties contributing to the advice.

\subsection{Experimental Procedure} 

The field experiment was conducted in person, with the assistance of trained research assistants who guided customers through the process. These research assistants approached and recruited 137 customers who visited two bank branches between October 4 and November 26, 2021. All customers were given a fixed participation fee of ten Euros. We also endowed the customers with 100 Euros for each of the ten loan investment opportunities, so investment decisions in one loan are independent of the others. Customers had two chances to invest in each loan, once before and once after receiving investment recommendations. They were told that a minimum of 10\% of all the participants in this study would be lottery winners for whom one of the twenty investment decisions (ten before and ten after receiving advice) was randomly drawn to realize the actual final investment outcome. 

Following the explanation of the potential payoffs, we asked customers to make investment decisions and risk assessments in seven categories (ranging from extremely low to extremely high) for the ten loan investment opportunities given the observed loan characteristics without any advice. After making the initial investment decisions, the customers learned they would receive advice and needed to make another round of decisions. Using a between-subjects design, we randomly assigned advice sources (AI-only, Human-AI, and Human-only). Table \ref{tab:field_source_stimuli} in the Online Appendix presents the description of the advisor customers received. Customers then made the second (final) set of investment decisions and risk assessments on the ten investment opportunities with the advice provided in addition to the loan characteristics (see an example of the interface in Figure \ref{fig:field_interface}). Finally, customers also provided information about their age and risk-taking tendencies by indicating how much they were willing to take risks on a scale from zero to ten \citep{dohmen2012intergenerational}. We summarize the main procedure of the field experiment in the right panel (Stage 2) of Figure \ref{fig:field_experimental_design}.

Customers took about seventeen minutes to complete the study and earned an average payoff of 15.72 Euros from their investments, which translates to an approximate hourly wage of 90 Euros. The average bonus among the 20 lottery-winning participants was 107.65 Euros.

\subsection{Data}

From the initial sample of 137 customers, we excluded seven due to poor-quality responses characterized by straight-lining (one from AI-only, three from Human-AI, and three from Human-only conditions), i.e., always inattentively making the same decision, in both initial risk assessments and investment decisions across all loans \citep{curran2016methods, meade2012identifying}.%%%
%%%
%%%
\footnote{We conducted a robustness check on the treatment effect of the Human-AI condition by including those straight-liners. The results are shown in Table \ref{table:field_main_robustness} Column (3).}
%%%
Our final analysis sample comprises 130 customers, distributed across the experimental conditions as follows: 45 in the AI-only condition, 42 in the Human-AI condition, and 43 in the Human-only condition. Since each customer was asked to make investment decisions for ten personal loans, and one customer in the AI-only condition did not make an initial decision for one loan, we obtained a customer-loan-level panel dataset consisting of 1,299 observations. We present an overview of the customer-loan-level analyses sample size for the field experiment in Table \ref{tab:sample}. As we focus on estimating the impact of Human-AI (vs. AI) on customers' reliance on advice, the main analyses involve 869 observations, as shown in Table \ref{tab:sample}.

Each observation in this dataset includes variables at multiple levels of analysis. At the individual customer-loan level, we have data on initial and final investment decisions, as well as initial and final risk assessments. Treatment-loan level variables, which are consistent across all customers within a specific treatment group, include advice on investment decisions and risk assessments. Additionally, loan-level variables, which are invariant across customers, encompass attributes such as the loan's targeted amount, term, annual percentage rate (APR), etc. Finally, customer-level variables, which remain constant across different loans for each customer, include treatment assignment, age, and risk-taking propensity.

\subsubsection{Dependent Variables}\label{dv}

We assess the persuasive efficacy of human-AI collaborative advice relative to AI-only advice by examining customers' reliance on the advice provided. Drawing on the advice-taking literature \citep{bonaccio2006advice}, we quantify a customer's degree of advice-taking using a dummy variable that indicates the alignment of their final investment decision with the advisor's recommendation, denoted as $FinalAlign$. This alignment measure serves as our primary dependent variable because the investment decisions are incentivized.

As a complementary measure of advice-taking, we also consider customers' final risk assessments, which are rated on the aforementioned seven-point scale. We measure reliance on advice by the absolute difference between the customer's final risk assessment and the observed assessment, denoted as $GapFinalRiskAssess$. However, given that the risk assessment was not incentivized and does not have a clear ``ground truth," we consider $GapFinalRiskAssess$ as a supplementary measure of advice-taking for robustness checks.%%%
%%%
%%%
\footnote{Weight on advice (WOA) is an alternative measure of the degree of advice-taking, which is only used for judgment tasks on a continuous scale (e.g., predicting a math test score ranging from 0 to 100) due to its truncation and removal of observations when the initial judgment aligns with the advice. Given the discrete nature of our decision variables, we follow the recommendation of \cite{bonaccio2006advice}, who also recognize the restriction of the WOA measure that does not consider a case as advice-taking when ``[t]he advice may have succeeded in convincing judges that their initial choices, quite possibly made under a great deal of uncertainty, were likely to be correct." We thus use the alignment of the judgment and decision to advice to measure the degree of advice-taking or the persuasive efficacy of the advice.}

\subsubsection{Control Variables}\label{control_var}

In our analyses, we aim to explain variations in bank customers' reliance on advice by controlling for several factors, including age, stated risk preferences, initial risk assessments, and initial investment decisions across different loan opportunities. To further examine the determinants of advice-taking behavior, we introduce additional variables as recommended by the literature. Prior research indicates that individuals are more likely to discount advice that deviates significantly from their initial opinions \citep{harvey1997taking, yaniv2004receiving}. To capture this phenomenon, we introduce two variables: the alignment (gap) between the initial and advised investment decision ($InitAlign$) and the gap between the initial and advised risk assessment ($GapInitRiskAssess$), which quantify the extent to which a customer's initial opinion diverges from the provided advice.

Financial investment decisions are often influenced by individuals' perceptions of risk, in addition to their inherent risk preferences \citep{weber1997perceived}. A substantial body of literature highlights the heterogeneity in how individuals translate risk assessments into financial decisions \citep[e.g.,][]{cooper1988entrepreneurs,weber2002domain}. Some individuals may not accurately assess the risks associated with investment opportunities due to tendencies such as denial of the existence of risks \citep{cooper1988entrepreneurs, march1987managerial}, or overconfidence in their ability to manage and mitigate these risks effectively \citep{strickland1966temporal,weber2002domain}. Others may exhibit inconsistencies in how their risk assessments influence their investment decisions \citep{maccrimmon1990characteristics}. To capture customers' sensitivity to risk perceptions in the context of financial investment decisions, we measure the range of their initial risk assessments ($RangeInitRisk$), anchored by their minimum risk assessment ($MinInitRisk$) across the ten initial investment decisions. Additionally, we assess the inconsistency in a customer's risk assessments as they relate to their investment decisions. We identify customers with inconsistent investment behaviors when there is an overlap between their lowest risk assessment for not investing and their highest risk assessment for investing, a phenomenon we term the ``investment decision gray zone,'' denoted as $DecisionGrayZone$. This variable quantifies the degree of inconsistency in customers' initial risk assessments. 

Table \ref{tab:vardef} in the Online Appendix summarizes the key variables we use for analyses and their descriptions. Table \ref{table:sumstat} presents the descriptive statistics of the variables.

\subsubsection{Randomization Check}

We conducted a randomization check on customer-level variables which indicates no statistically significant differences between the two focal experimental conditions (i.e., the Human-AI and the AI-only conditions) in terms of customers' age, risk-taking tendencies, sensitivity in risk assessments across various investment opportunities, or inconsistency in how risk assessments translate into investment decisions (see Table \ref{tab:balance_check_ind} in the Online Appendix). This suggests that the randomization was successful at the customer level. Additionally, we performed a randomization check on individual customer-loan level variables prior to exposure to the advice. The findings, detailed in Table \ref{tab:balance_check_ind_loan} Columns (1)-(2) in the Online Appendix, similarly reveal no statistically significant differences across the experimental conditions in initial investment decisions and risk assessments. 

Together, these results confirm the validity of our experimental randomization between our focal experimental conditions (i.e., the Human-AI and the AI-only conditions).\footnote{Considering Human-only condition as an additional control treatment benchmarking with the Human-AI condition, further randomization checks confirm no statistically significant differences on all the customer-loan level variables shown in Table \ref{tab:balance_check_ind_loan}, and all customer level variables shown in Table \ref{tab:balance_check_ind} except for the $RangeInitRisk$, which we control for in the relevant regression analyses.} To reduce potential estimation biases stemming from the variations in advice between the Human-AI and AI-only scenarios, we account for such differences in advice and how they diverge from an individual's initial risk assessment and investment decisions in later analyses.

\subsubsection{Model-free Evidence}

\begin{figure}[h!]
    \centering
    \footnotesize
    \caption{\label{fig:modelfree}Model-free evidence when the Human-AI and AI advisor offered the same investment advice}
     \centering
     \begin{subfigure}[b]{0.48\textwidth}
         \centering
         \scriptsize
         \caption{Alignment in Final Investment with Advice\label{fig: modelfree_invest}}
         \includegraphics[width=0.9\textwidth]{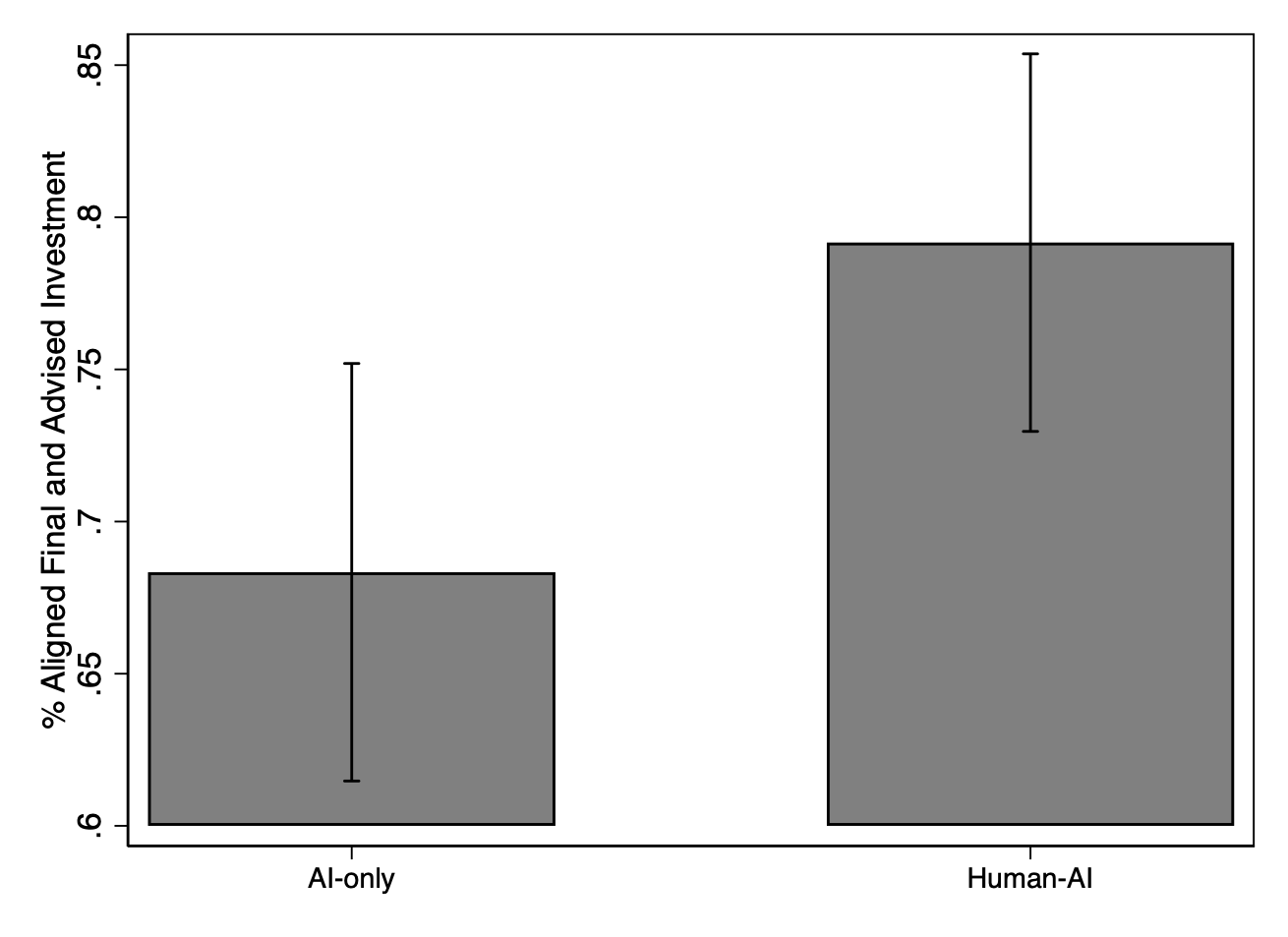}
     \end{subfigure}
     \hfill
     \begin{subfigure}[b]{0.48\textwidth}
         \centering
         \scriptsize
         \caption{Alignment in Final Risk Assessment with Advice\label{fig: modelfree_risk}}
         \includegraphics[width=0.9\textwidth]{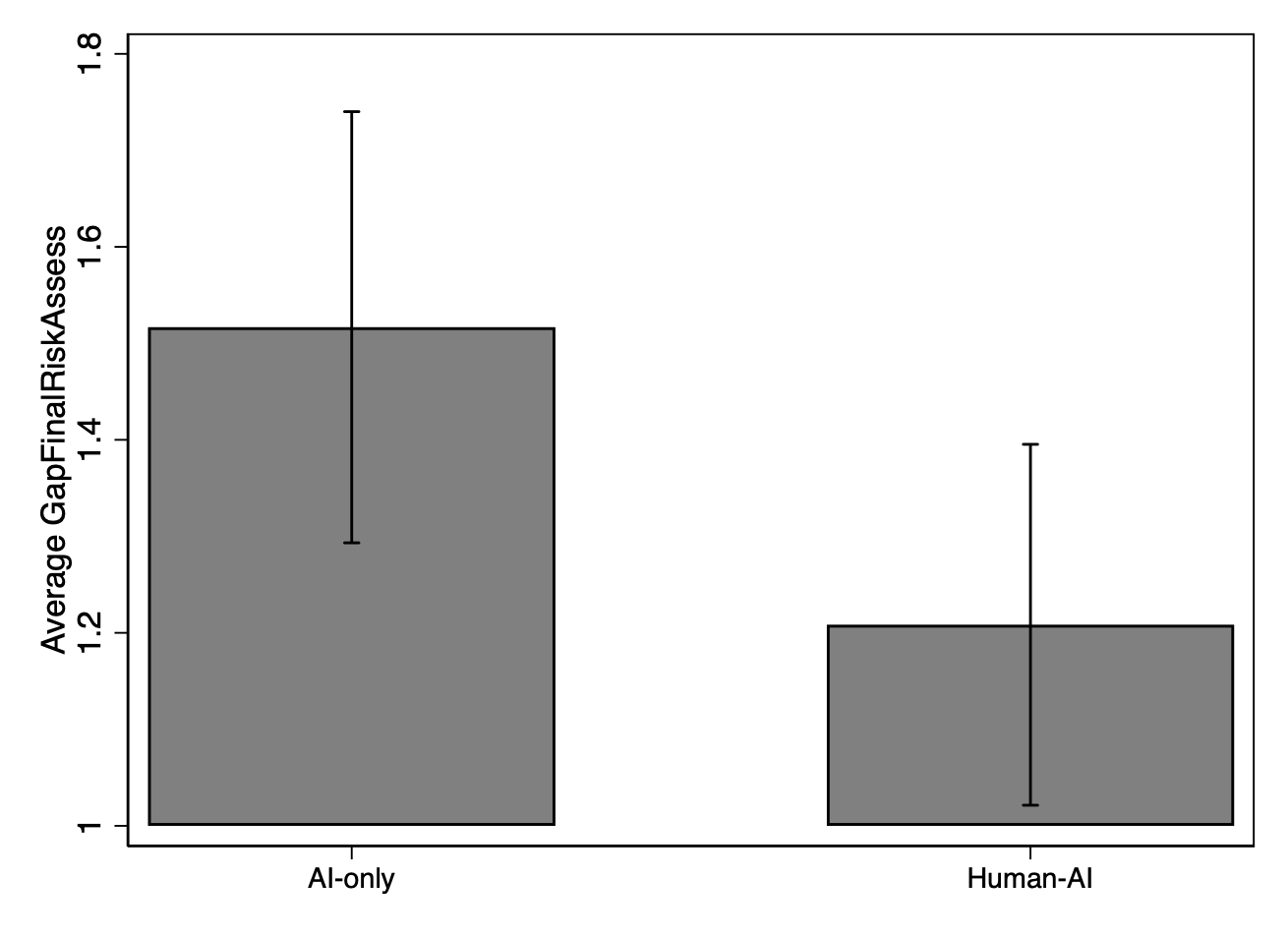}        
     \end{subfigure}    \\
    Note: Error bars indicate 95\% confidence intervals. 
\end{figure}

We first present model-free evidence on bank customers' reliance on different advice sources across experimental conditions. Since the advised investment recommendations and risk assessments differ for six out of ten loans between the Human-AI and AI-only conditions, isolating the pure treatment effect of having a human in the loop from differences induced by varying advice provided is challenging. For a fair comparison, we focus on the subsample of loans for which both the investment advice and risk assessments were consistent across conditions. As shown in Figure \ref{fig: modelfree_invest}, the percentage of aligned investment decisions is higher in the Human-AI condition ($M=0.792$, $SD=0.407$, $N=168$) than in the AI-only condition ($M=0.683$, $SD=0.466$, $N=180$), with this difference being %economically (16\%) and 
statistically significant ($p<0.05$; two-sample Student's t-test). Similarly, Figure \ref{fig: modelfree_risk} illustrates that customers exhibit greater alignment (20.4\%) in their final risk assessments with the advice in the Human-AI condition ($M=1.208$, $SD=0.095$, $N=168$) compared to the AI-only condition ($M=1.517$, $SD=0.113$, $N=180$). This difference is also statistically significant ($p<0.05$; two-sample Student's t-test). 

The preliminary model-free evidence suggests that bank customers follow human-AI collaborative advice to a greater extent than from an AI alone. However, we remain agnostic about the conclusiveness of the results, primarily due to the exclusion of data on six loans where investment advice differs in the Human-AI and AI-only conditions. We next formalize our investigation through regression analyses using the alignment in the final investments with the observed investment recommendation as our main dependent variable, given that it is incentivized and has implications in actual financial returns. This regression analysis allows us to control for various, possibly important co-variates and fixed-effects.

\subsection{Results}
In this section, we examine whether and why bank customers are inclined to align their investment decisions with the investment recommendations under the Human-AI condition compared to the AI-only control condition. We first present the average treatment effect of the Human-AI condition as the main results. Subsequently, we leverage the Human-only condition and subsamples of more and less uncertain investments to investigate the underlying mechanism.

\subsubsection{Main Results}

Equation (\ref{eq_specification}) outlines the estimation of the treatment effect of Human-AI (versus AI-only) on the alignment of customer $i$'s final investment decision with the investment recommendation received for loan $l$. Given our interest in directly estimating the average treatment effect of Human-AI versus AI-only in percentage points, we adopt linear regression using ordinary least squares (OLS) as our primary specification  \citep{angrist2009mostly,wooldridge2010econometric}. The linear specification is also less susceptible to biased estimates of the treatment effect due to the incidental parameter problem arising from the fixed time dimension of a panel using logistic regression \citep{greene2004fixed}. We will later present the robustness checks by using non-linear regression methods such as logistic regression.%%%
%%%
\footnote{Note that 99.88\% of the predicted probabilities stay in the boundary of 0 and 1 for our focal sample of 869 observations involving Human-AI and AI-only conditions, mitigating concerns about the applicability of OLS.}
%%%

The primary independent variable of interest is the dummy variable $HumanAI_{i}$, coded as 1 for the Human-AI condition and 0 for the AI-only condition. A positive (negative) estimate of $\beta_1$ would indicate that the alignment of the final investment decision with the advice is stronger (weaker) under the Human-AI condition compared to the AI-only baseline. %Conversely, a negative $\beta_1$ would suggest weaker alignment in the Human-AI condition. 
In our regression analyses, we control for customer-loan level variables ($X_{i,l}$), e.g., the alignment of the investment recommendation with the initial investment decision, and customer-level variables ($X_i$), e.g., customers' gender (see Section \ref{control_var} for an overview of controls). We include day-of-visit ($Date$) fixed effects, $\eta_i^t$, to capture time-variant factors that may influence investment decisions. We also include loan-level fixed effects to account for time-invariant characteristics specific to each loan and advice fixed effects contingent on loans to control for potential confounding between advice and treatment differences across six loans where the advice varies. This specification ultimately incorporates loan-by-advice fixed effects $\phi_{l}^{i}$, controlling for the 16 different loan features alongside the advice (10 loans, including 6 with varying $AdviceInvest$ and $AdviceRiskAssess$).%%%
%%%
\footnote{An alternative identification strategy involves a Difference-in-Differences (DiD) estimation to assess changes in $InvestAlign$ after receiving Human-AI advice compared to AI-only advice. However, concerns arise regarding the parallel trend assumption since differing advice on six of the ten loans could affect $InvestAlign$ outcomes before any advice. This variation is due to the experimental design centered on realistic advice. In contrast, our primary identification strategy in Equation \ref{eq_specification} allows for direct control over variations in advice through covariates. Nevertheless, DiD estimation on four loans with consistent advice across Human-AI and AI-only conditions, which avoids violations of the parallel trend assumption, produces results that align with our main findings.}
%%%
For ease of interpretation, we apply mean-centering for all continuous customer-loan level and customer-level variables.

%%%%%%%%%%%%%%%%%%%%%%%%%%%%%%%%%%%%%%%%%%%%%
%%%%%%%%%%%%%%%%%%%%%%%%%%%%%%%%%%%%%%%%%%%%%
\begin{equation} \label{eq_specification}
        FinalAlign_{i,l} = \beta_0+\beta_1 HumanAI_i + \gamma_1 X_{i,l}+ \gamma_2 X_{i} + \phi_{l}^i+ \eta_i^t +\epsilon_{i,l}
\end{equation}
%%%%%%%%%%%%%%%%%%%%%%%%%%%%%%%%%%%%%%%%%%%%%
%%%%%%%%%%%%%%%%%%%%%%%%%%%%%%%%%%%%%%%%%%%%%

Column (1) of Table \ref{table:field_main} reports the OLS regression results. Column (2) of Table \ref{table:field_main} further shows the results that cluster the robust standard errors at the individual customer level, thus allowing for any arbitrary error correlations within each customer. The coefficient estimates of $\beta_1$ are positive and statistically significant. We find no conclusive interpretation regarding other control variables, except that customers’ final investment alignment with advice decreases when their initial investment decision deviates from the advice and as the advisor’s age increases. The results shown in Table \ref{table:field_main} collectively demonstrate that our findings in the model-free evidence remain robust, even after including observations on six loans with differing investment recommendations and incorporating a variety of control variables. We gauge the economic significance of the estimated effect according to the coefficient estimate of $\beta_1$ shown in Column (2) of Table 1, concerning instances where the AI-only and Human-AI conditions provided the same or differing advice. This is due to the realistic consideration that human involvement may have changed the AI advice, such as in our setting. The estimated coefficient indicates an 15.5 percentage point increase as the average treatment effect in bank customers' final investment alignment with advice when AI-based investment recommendations involve a banker. 

%%%%%%%%%%%%%%%%%%%%%%%%%%%%%%%%%%%%%%%%%%%%%
%%%%%%%%%%%%%%%%%%%%%%%%%%%%%%%%%%%%%%%%%%%%%
\begin{table}[h!]
        \centering
        \caption{\label{table:field_main}Bank customers align their final investments with advice to a greater extent under the Human-AI than the AI-only condition -- OLS regression results.}
        \begin{threeparttable}
            \small
            {
\def\sym#1{\ifmmode^{#1}\else\(^{#1}\)\fi}
\begin{tabular}{l*{2}{c}}
\hline\hline
& \multicolumn{2}{c}{DV: FinalAlign}\\
\cline{2-3}
                              &\multicolumn{1}{c}{(1)}          &\multicolumn{1}{c}{(2)}          \\
\hline
Human-AI (vs. AI-only)        &               0.155\sym{***} &               0.155\sym{***} \\
                              &             (0.048)          &             (0.046)          \\
InitInvest                    &              -0.004          &              -0.004          \\
                              &             (0.037)          &             (0.030)          \\
InitAlign                     &               0.269\sym{****}&               0.269\sym{****}\\
                              &             (0.034)          &             (0.038)          \\
InitRiskAssess                &              -0.009          &              -0.009          \\
                              &             (0.011)          &             (0.010)          \\
GapInitRiskAssess             &              -0.008          &              -0.008          \\
                              &             (0.013)          &             (0.014)          \\
Age                           &              -0.002\sym{**}  &              -0.002\sym{***} \\
                              &             (0.001)          &             (0.001)          \\
RiskPref                      &              -0.011          &              -0.011          \\
                              &             (0.008)          &             (0.007)          \\
DecisionGrayzoneSize          &              -0.018          &              -0.018          \\
                              &             (0.010)          &             (0.012)          \\
MinInitRisk                   &               0.021          &               0.021          \\
                              &             (0.031)          &             (0.031)          \\
Constant                      &               0.453\sym{****}&               0.453\sym{****}\\
                              &             (0.036)          &             (0.038)          \\
\hline
Observations                  &                 869          &                 869          \\
\(R^{2}\)                     &               0.244          &               0.244          \\
Loan $\times$ Advice fixed-effects&                 Yes          &                 Yes          \\
Date fixed-effects            &                 Yes          &                 Yes          \\
Clustered standard errors     &                  No          &                 Yes          \\
\hline\hline
\end{tabular}
}

            \begin{tablenotes}[flushleft]
                \item\leavevmode\kern-\scriptspace\kern-\labelsep\textit{Notes}: Standard errors are reported in the parentheses; **$p<0.05$, ***$p<0.01$, ****$p<0.001$.       
            \end{tablenotes}
        \end{threeparttable}
\end{table}
%%%%%%%%%%%%%%%%%%%%%%%%%%%%%%%%%%%%%%%%%%%%%
%%%%%%%%%%%%%%%%%%%%%%%%%%%%%%%%%%%%%%%%%%%%%

We performed several robustness checks and report the results in Table \ref{table:field_main_robustness} in the Online Appendix. First, Column (1) of Table \ref{table:field_main_robustness} presents qualitative similar results as shown in Table \ref{table:field_main} using a non-linear logistic model specification specification. 
Since we are interested in estimating an average treatment effect of the Human-AI (vs. AI-only) condition, we, however, continue to focus on the linear specification in the rest of the analyses. 
Second, Column (2) of Table \ref{table:field_main_robustness} reports the OLS regression results when we include the fixed effects of the branch to which the customer belongs. We obtain similar magnitudes of the coefficient estimates of $\beta_1$ compared to that shown in Column (2) of Table \ref{table:field_main}, which confirms the assessment of the branch managers that the customers associated with the two branches are quite similar as they reside in two comparable towns at close distance. Third, Column (3) of Table \ref{table:field_main_robustness} reports the OLS regression results when we include customers who provided poor quality responses (straight-lining responses in risk assessment and investment decisions) while including a dummy variable indicating their response pattern. The coefficient estimate of $\beta_1$ is similar to that shown in Column (2) of Table \ref{table:field_main}. The significantly negative coefficient estimate of the dummy variable of $Straightliner$ indicates that customers who provide straight-lining responses are less likely to align their final investment decision to advice. These results are consistent with prior studies that straight-liners generally add noise to the detected effect due to low involvement \citep{curran2016methods,meade2012identifying} and thus, we exclude from the rest of the analyses. Fourth, Column (4) of Table \ref{table:field_main_robustness} reports the OLS regression results when we exclude loans with different investment advice. The coefficient estimate of $\beta_1$ is similar to that shown in Column (2) of Table \ref{table:field_main}, confirming that the model-free evidence shown in Figure \ref{fig: modelfree_invest} is robust after controlling for customer and customer-loan level variations, such as initial risk assessment and investment tendency. Last, Column (5) of Table \ref{table:field_main_robustness} reports the OLS regression results using our alternative dependent variable $GapFinalRiskAssess$ with the same specification shown in Equation \ref{eq_specification}. This regression provides insights into whether the human in the loop also affects advice-taking in the realm of customers' assessments of loan risks. Corroborating our previous results, we find that the estimate of $\beta_1$ is significantly negative, indicating that customers under the Human-AI (vs. AI-only) condition deviate less from the advised risk assessment (i.e., greater alignment in the final risk assessment with advice).

The consistent results presented in Tables \ref{table:field_main} and \ref{table:field_main_robustness} reinforce our confidence in the conclusion that customers are more likely to follow AI-based recommendations when a banker is involved in the advice generation process.

\subsubsection{First Evidence on Underlying Mechanism}\label{theory}

We next provide preliminary evidence on the underlying mechanism as to why bank customers increase their investment alignment with advice in the Human-AI compared to AI-only condition. In particular, we begin our investigation by leveraging the Human-only condition to examine whether a specific central route element, the banker-AI complementarity, drives the bank customers' greater reliance on human-AI collaborative advice. Subsequently, we advance our examination by investigating whether the positive effect of having a human in the loop on advice-taking stems more from the central or peripheral route by investigating the moderating role of decision uncertainty per our discussion in Section \ref{theory}.

\paragraph{The Role of Human.}

We leverage the additional control treatment condition, ``Human-only," to start exploring the mechanism through which a banker's involvement in AI-based advice influences customer decision-making. As outlined in Section \ref{field_exp_design}, we conjecture that if the alignment of investment decisions with advice is lower in the Human-only condition compared to the Human-AI condition, this would suggest that the main treatment effects are driven by customers being persuaded through the expected advantages of having multiple entities involved in the advice production process, which corresponds to the central route of persuasion. Conversely, if there is no significant difference in the alignment of investments with advice between the Human-only and Human-AI conditions, it would indicate that the main treatment effect of Human-AI (vs. AI-only) is more nuanced. This scenario suggests that the effect may not solely arise from the cognitive perception of multiple entities contributing to the advice, but could involve peripheral factors like customers' emotional responses to the banker's involvement.

%%%%%%%%%%%%%%%%%%%%%%%%%%%%%%%%%%%%%%%%%%%%%
%%%%%%%%%%%%%%%%%%%%%%%%%%%%%%%%%%%%%%%%%%%%%
\begin{table}[h!]
    \centering
    \caption{Bank customers do not reduce their alignment in the final investments with advice under the Human-only than the Human-AI condition -- OLS regression results.\label{table:field_human}}
        \begin{threeparttable}
            \small
            {
\def\sym#1{\ifmmode^{#1}\else\(^{#1}\)\fi}
\begin{tabular}{l*{1}{c}}
\hline\hline
                              &\multicolumn{1}{c}{DV: FinalAlign}          \\
\hline
Human-AI (vs. Human-only)     &              -0.054          \\
                              &             (0.037)          \\
Constant                      &               0.500\sym{****}\\
                              &             (0.034)          \\
\hline
Observations                  &                 849          \\
\(R^{2}\)                     &               0.392          \\
Loan fixed-effects            &                 Yes          \\
Date fixed-effects            &                 Yes          \\
Controls                      &                 Yes          \\
\hline\hline
\end{tabular}
}

            \begin{tablenotes}[flushleft]
                \footnotesize
                \item\leavevmode\kern-\scriptspace\kern-\labelsep\textit{Notes}: Standard errors clustered at the customer level are shown in the parentheses; **$p<0.05$, ***$p<0.01$, ****$p<0.001$.       
            \end{tablenotes}
        \end{threeparttable}
    
\end{table}
%%%%%%%%%%%%%%%%%%%%%%%%%%%%%%%%%%%%%%%%%%%%%
%%%%%%%%%%%%%%%%%%%%%%%%%%%%%%%%%%%%%%%%%%%%%

We compare customers' alignment in their final investments with advice under the Human-AI and the Human-only conditions by applying the same specification as shown in Column (2) of Table \ref{table:field_main} with one key modification of using the Human-only condition as the reference category.\footnote{Since advice is identical for the same loan across the Human-AI and Human-only conditions, we only include loan fixed effects} Table \ref{table:field_human} presents the OLS regression results, suggesting that customers' alignment in their final investment decisions with advice does not differ between the Human-AI and the Human-only conditions. This observation challenges the notion that the positive average treatment effect of the Human-AI condition relative to the AI-only condition (as shown in Table \ref{table:field_main}) is driven by the perception of a synergistic ``two-advisors" effect or banker-AI complementarity.

\paragraph{The Role of Uncertainty.}

We next investigate whether the central route (e.g., positive beliefs about human involvement) or the peripheral route (e.g., enhanced decision comfort engendered by social cues associated with human involvement) has a relatively more substantial influence on the heightened persuasive efficacy of human-AI collaborative advice through a heterogeneity analysis. 

Prior literature suggests that decision uncertainty may increase individuals' perceived decision complexity and thus induces peripheral instead of central route of information processing upon decision making \citep{tversky1974judgment}. As such, we expect peripheral factors to play a dominant role in explaining why bank customers are more likely to rely on human-AI collaborative advice if this positive effect is most evident when they experience greater decision uncertainty. On the other hand, we do not expect uncertainty to differentiate bank customers' reliance on the human-AI collaborative from pure AI advice.

We analyze whether treatment effects vary based on the level of investment uncertainty. Although we do not have a direct measure of bank customers' decision uncertainty, we use risk assessment by the advisor as a proxy, where elevated risk assessments suggest increased investment uncertainty.%%%
%%%
%%%
\footnote{Note: Although risk assessments and decision uncertainty are conceptually distinct -- with risk assessments involving judgments about the probability of an adverse event, and uncertainty reflecting a lack of certainty about such probabilities -- the two constructs are usually positively correlated \citep[e.g.,][]{slovic1987perception}. We also used bank customers' initial risk assessments (before receiving advice) as a proxy for decision uncertainty. The results are consistent and available upon request from the authors. We prefer using the advisor's risk assessment because a median split of the sample is more balanced, considering our loan selection is relatively evenly distributed in the risk assessment and, therefore, less sensitive to the median cutoff.}
%%%
%%%
For ease of interpretation, we divide the sample into two groups based on the advised risk assessment,  where we define a relatively high-risk assessment level as greater than the median level of 4 (labeled as moderate risk) out of 7 levels (ranging from extremely low to extremely high risk). Table \ref{table:field_moderation} presents the results using the same specification as in Column (2) of Table \ref{table:field_main}. The results in Columns (1) and (2) of Table \ref{table:field_moderation} indicate that customers exhibit greater positive alignment with the advice in the Human-AI (vs. AI-only) condition for more risky loans. However, this effect is not observed for less risky loans. Although not conclusive by any means, these findings support the notion that human involvement in AI-based recommendations serves as a peripheral cue, which ultimately makes customers follow the advice more. We test the potential underlying mechanism more directly in our controlled online experiment, allowing us to provide more conclusive evidence on why customers are more aligned with advice when human advisors are in the loop of AI-based investment advice production.

\begin{table}
    \centering
    \caption{\label{table:field_moderation}Bank customers align their final investments with advice to a greater extent under the Human-AI than the AI-only condition, especially when making more risky investments -- OLS regression results.}
    \begin{threeparttable}
        \footnotesize
        {
\def\sym#1{\ifmmode^{#1}\else\(^{#1}\)\fi}
\begin{tabular}{l*{2}{c}}
\hline\hline
                                 &   \multicolumn{2}{c}{DV: FinalAlign}                            \\
                                 \cline{2-3}
                                 &\multicolumn{1}{c}{(1)} &\multicolumn{1}{c}{(2)} \\
                                 &\multicolumn{1}{c}{More Risky Investments}&\multicolumn{1}{c}{Less Risky Investments}\\
\hline
Human-AI (vs. AI-only)           &               0.213****&               0.094    \\
                                 &             (0.060)    &             (0.062)    \\
Constant                         &               0.382****&               0.502****\\
                                 &             (0.075)    &             (0.045)    \\
\hline
Observations                     &                 432    &                 437    \\
\(R^{2}\)                        &               0.322    &               0.225    \\
Loan $\times$ Advice fixed-effects&                 Yes    &                 Yes    \\
Date fixed-effects                &                 Yes    &                 Yes    \\
Controls                         &                 Yes    &                 Yes    \\
\hline\hline
\end{tabular}
}

        \begin{tablenotes}[flushleft]
            \footnotesize
            \item\leavevmode\kern-\scriptspace\kern-\labelsep\textit{Note}: Standard errors clustered at the customer level are shown in the parentheses; **$p<0.05$, ***$p<0.01$, ****$p<0.001$.  
        \end{tablenotes}
    \end{threeparttable}    
\end{table}

\subsection{Discussion of field study}    

The results of our field experiment reveal that involving a human in an AI-driven financial advisory process (i) changes the nature of the final advice in the upstream production process, and (ii) increases the extent of customers' following the final advice downstream.

From a theoretical perspective, our findings in the field experiment underscore the critical role of human presence in AI-driven advisory systems, indicating that the human component acts as a facilitator of advice acceptance. Moreover, the lack of variation in customer reliance on the human-AI collaborative advice when not knowing about the AI's involvement suggests that the positive response to human participation in the advisory process does not simply stem from an expected advantage of two complementary entities involved in the advice production. We find that the treatment effect is driven by scenarios where customers perceive relatively high uncertainty, an observation that aligns with the ELM, suggesting that the peripheral route of persuasion is more influential under conditions of high uncertainty. Notably, in our specific setting where human involvement does not adversely affect AI advice quality, the higher inclination to adhere to received advice is materially beneficial to customers. We measure the economic consequences of customers' higher reliance on advice from human-AI collaboration compared to a pure AI by analyzing differences in consumer surplus via their final payoffs after receiving advice. Column (1) of Table \ref{table:field_payoff} shows the OLS regression results of the final payoff (log-transformed due to right-skewness) when using the same specification as in Column (2) of Table \ref{table:field_main}. As we find no difference in the investment payoffs before receiving advice (see results shown in Table \ref{tab:balance_check_ind_loan}), the results shown in Column (1) of Table \ref{table:field_payoff} suggest an average increase of 44.92\% in the final payoff for the bank customers under the Human-AI than the AI-only condition across all loans. Additionally, Columns (2) and (3) of Table \ref{table:field_payoff} present the OLS regression results on the same subsamples of more versus less risky investments as shown in Table \ref{table:field_moderation}. These results suggest that the increase in customers' payoff when utilizing the human-AI collaborative advice is mainly driven by greater alignment with the advice when making investments under uncertainty.\footnote{The correlation between risk assessment and APR is 0.82 in our data ($p<0.001$).} However, we caution against generalizing these results to contexts where AI predictions are inherently unreliable. In such cases, consumers may be disadvantaged by over-relying on human-AI collaborative advice due to misplaced trust rather than correctly discounting suboptimal advice.

\begin{table}
    \centering
    \caption{\label{table:field_payoff}Bank customers earn higher payoffs after receiving advice under the Human-AI than the AI-only conditions, especially when making more risky investments -- OLS regression results}
    \begin{threeparttable}
        \footnotesize
        {
\def\sym#1{\ifmmode^{#1}\else\(^{#1}\)\fi}
\begin{tabular}{l*{3}{c}}
\hline\hline
& \multicolumn{3}{c}{DV:$\log$(FinalPayoff+1)}\\
\cline{2-4}

                                 &        Whole Sample    &More Risky Investments    &Less Risky Investments    \\
                                 &\multicolumn{1}{c}{(1)}    &\multicolumn{1}{c}{(2)}    &\multicolumn{1}{c}{(3)}    \\
\hline
Human-AI (vs. AI-only)           &               0.371*** &               0.791*** &              -0.014    \\
                                 &             (0.139)    &             (0.288)    &             (0.055)    \\
Constant                         &               4.195****&               4.002****&               4.526****\\
                                 &             (0.098)    &             (0.262)    &             (0.052)    \\
\hline
Observations                     &                 869    &                 432    &                 437    \\
\(R^{2}\)                        &               0.415    &               0.358    &               0.594    \\
Loan $\times$ Advice fixed-effects&                 Yes    &                 Yes    &                 Yes    \\
Date fixed-effects               &                 Yes    &                 Yes    &                 Yes    \\
Controls                         &                 Yes    &                 Yes    &                 Yes    \\

\hline\hline
\end{tabular}
}

        \begin{tablenotes}[flushleft]
            \footnotesize
            \item\leavevmode\kern-\scriptspace\kern-\labelsep\textit{Note}: Standard errors clustered at the customer level are shown in the parentheses; **$p<0.05$, ***$p<0.01$, ****$p<0.01$.
        \end{tablenotes}
    \end{threeparttable}
    
\end{table}

\section{Online Controlled Experiment}\label{lab}

We conducted a controlled and incentivized online experiment to uncover the underlying mechanism that may drive the main findings from the field experiment. We aimed to find direct process evidence for the central or peripheral persuasion routes and explore which persuasive element(s) induce individuals' greater reliance on human-AI collaborative recommendations. We advance our understanding of the underlying mechanism by testing both central and peripheral route elements that could account for our findings as discussed in Section \ref{persuasion}. We formalize our test of the underlying mechanism by conducting a moderated mediation analysis \citep{li2022recommender} where directly measured decision uncertainty positively moderates individuals' perception about human involvement in the AI-based advice via central route elements (cognitive trust or satisfaction in advisor's prediction performance) or peripheral route elements (emotional trust, relinquishment of decision autonomy, and shared decision responsibility) that mediate one's alignment in the final investments to advice.

\subsection{Experimental Setup}

\subsubsection{Participants}
We recruited 100 German-speaking participants (36.36\% female, $M_{age}=32.6$) from July 30 to August 2, 2024, via the Prolific platform. The online experiment was supposed to be as close as possible to the field experiment, which we conducted in cooperation with a savings bank in Germany. We planned to exclude the following participants from the sample of analyses: those who declined to consent, failed to pass the attention check, and had poor-quality responses (e.g., straight-liners).

\subsubsection{Experimental Design}

As we can demand explanations for decisions from participants recruited from an online panel with less time constraint than in the field experiment, we directly measured possible constructs that could contribute to individuals' greater reliance on human-AI collaborative recommendations. We dropped the Human-only condition, which we used as a control treatment in the field experiment. Besides measuring the decision uncertainty and the potential persuasive elements, we also measured participants' prior belief in a banker's, banker-AI's, and AI's investment advice accuracy on a continuous scale from zero to 100\% before informing them about their actual advisor. The prior belief measures enable us to test whether the prior belief in AI and human banker's prediction accuracy differs from the subsequently communicated advice accuracy of the advice source (always 70\%), which could have contributed to participants' lower reliance on AI advice and greater reliance on human banker's advice shown in the field experiment. Additionally, the discrepancy between the prior belief in a banker-AI pair versus an AI's prediction performance could potentially contribute to the differences in the reliance on advice observed in the field experiment, which we can also test in this controlled online experiment. 

Another purpose of this online experiment is to show the robustness of our field experiment findings by removing potentially confounding factors to the treatment in the field experiment. This involves: 1) using the same loans as investment opportunities presented in identical decision order as in the field experiment, 2) providing identical advice on risk assessments and investment recommendations across the experimental conditions by using the human-AI collaborative advice shown in Table \ref{tab:loan_advice_details}, 3) using a similar description for different advisors, and 4) informing identical advice accuracy of the advisor (AI or human-AI) at 70\% across conditions after eliciting the prior beliefs in the advisors' prediction performance.

\subsubsection{Experimental Procedure}
We told all participants they would make investment decisions on ten personal loans, followed by a short questionnaire in exchange for a fixed participation fee of 2£ plus a potential bonus. We first explained the incentive mechanism. Each participant was endowed with 500 points for each investment opportunity, with each point equivalent to 0.15 pence, i.e., the 500 points were worth 0.75£. Identical to the field experiment, each participant made an investment decision and risk assessment for each personal loan, once before and once after receiving advice. One of the twenty investment decisions (ten before and ten after receiving advice) was randomly selected to realize the final payoff. We rewarded participants a bonus for the difference in the final payoffs and the fixed participation fee if the final payoff exceeds the fixed participation fee of 2£.

After explaining the potential payoff, we asked the participants to provide their initial risk assessments and investment decisions, as well as their confidence in their decisions on a scale from one to five for each of the ten loans to measure their decision (un)certainty \citep{bonaccio2006advice,dietvorst2015algorithm}. Following the initial evaluations, we asked all participants to give their (prior) belief in a potential AI advisor, human advisor (banker), and human advisor using AI (banker-AI)'s prediction accuracy ranging from zero to 100\%  \citep{dietvorst2015algorithm}, presented in a random order. We explained that an accurate prediction means the advisor recommends investing when a loan does not default and not investing when it does. We then randomly assigned participants to one of the two experimental conditions (i.e., AI-only and Human-AI) with a description of the advisor. Table \ref{tab:lab_source_stimuli} shows the English translation of the description. The original German version of the advisor description is of similar text length (119 and 123 words for the AI-only and the Human-AI conditions, respectively). All participants were informed ``The recommendation of the \{the artificial intelligence / bank advisor who worked with artificial intelligence\} is correct in about 70\% of cases" according to their assigned experimental condition before proceeding to provide the final risk assessments, investment decisions, and confidence in their decisions. 

Before finishing the study, we asked all participants to indicate their assigned advisor from a choice of three options (banker, AI, or banker assisted by an AI) as an attention check. We also asked participants to indicate their perception of the decision responsibility (choice option among the participant, the advisor, or both the advisor and the participant) \citep{steffel2016passing} and their perceived decision autonomy \citep{dalal2010types}, cognitive trust (measured by three items, Cronbach $\alpha = 0.41$) and emotional trust (measured by three items, Cronbach $\alpha = 0.78$) in their advisor \citep{komiak2006effects,mcknight2002developing}, disappointment in the advisor's prediction accuracy \citep[adapted from][]{dietvorst2015algorithm}. We present the item details in Table \ref{table:perceptual_measure} in the Online Appendix. Due to the low reliability score on the three items related to cognitive trust, we dropped the third item (the least correlated) from calculating the construct also because it is considered ambiguous in its German translation by an independent German-speaking research assistant. We also dropped the first item from the emotional trust construct for the same reason. The correlations between the two items in constructing the cognitive and emotional trust index are 0.50 and 0.72 ($ps<0.05$), respectively. We use the average rating of related items as a composite index to measure cognitive and emotional trust. We code perceived shared decision responsibility as one when a participant does not solely contribute the decision responsibility to herself and zero otherwise, denoted as $AdvisorResp$. Participants also indicated their age, gender, nationality, and risk preferences \citep{dohmen2012intergenerational}. Table \ref{table:sumstat_lab} in the Online Appendix shows the descriptive statistics of the variables. 

Since the sample might be inherently different in their financial investment experience compared to the bank customer in the field setting, we also asked the participants to indicate their investment experience ranging from one to seven \citep{mishra2015study}.

\subsection{Results}

\subsubsection{Replication of Field Experiment Results}

Out of the 100 participants, we excluded thirteen from the analysis. Among the thirteen participants, ten (three from the Human-AI condition and seven from the AI-only condition) did not pass the attention check, while three (two from the Human-AI condition and one from the AI-only condition) offered low-quality responses (e.g., straight-liners). Among the 87 participants, 44 were assigned to the AI-only and 43 to the Human-AI conditions. Since each participant invested in ten personal loans, we obtained individual-loan-level panel data with 870 observations. 

To ensure the comparability of the results from the online laboratory experiment and the field experiment, we first assess the samples from our field and online experiments with respect to control variables and first-round decisions before receiving advice. Participants in the laboratory experiment are significantly younger than the bank customers in our field experiment ($p<0.01$, two-sample Student's t-test), likely due to the specific demographic of those visiting the savings bank in Germany. However, we find no statistically significant differences between the field and laboratory experiment participants regarding initial risk assessments ($p=0.31$, two-sample Student's t-test) and investment decisions ($p=0.17$, Pearson's Chi-squared test). Aside from age, which we control for in all analyses, there is no evidence suggesting that participants in the laboratory experiment differ from bank customers in their investment behaviors before receiving advice.

The critical question is whether our controlled online experiment replicates the results observed in the field experiment. Initial model-free evidence suggests that it does: participants in the online experiment are more likely to align their final investment decisions with advice in the Human-AI condition ($M=0.71$, $SD=0.46$, $N=430$) compared to the AI-only condition ($M=0.65$, $SD=0.48$, $N=440$). This difference, is both economically meaningful (+9.2\%) and approaches statistical significance even in this model-free testing where we do not control for co-variates and clustering of standard errors (two-sample Student's t-test, $p=0.08$).

To formalize the analysis, we adopt the specification outlined in Equation \ref{eq_specification} with two adjustments to the exact one used for the field experiment (i.e., Column (2) of Table \ref{table:field_main}) to assess the alignment of participants' final investments with the advice given under the AI-only and Human-AI conditions. First, we exclude the advised investment and risk assessment variables because loan-level fixed effects absorb their variation in the laboratory setting, where advice was uniformly distributed across conditions. Second, we include the same individual-level control variables as in the field experiment, with additional variables for participants' gender %, financial investment experience (coded as one if rated four or higher on a seven-point scale, and zero otherwise), 
and prior beliefs in the prediction accuracy of banker-AI versus AI, which were not available in the field setting. All continuous variables are mean-centered, maintaining consistency with the field experiment methodology.

Table \ref{table:lab_main} Column (1) presents the OLS regression results, showing that participants align their final investments with advice more in the Human-AI condition than in the AI-only condition, with the estimate of $\beta_1$ being not significantly different from that of the field experiment shown in Column (2) of Table \ref{table:field_main} ($p=0.14$, two-tailed Z-test). We note that the greater alignment in the final investment to the advice is present despite controlling for participants' prior belief in the prediction accuracy of banker-AI and AI advisor, whose coefficient estimates are not significantly different from zero (see Column (1) of Table \ref{table:lab_main}). We further exclude participants' prior belief in the prediction accuracy of the banker-AI and AI advisor from the analyses, and the results are shown in Column (2) of Table \ref{table:lab_main}. The coefficient estimate of $\beta_1$ shown in Column (2) of Table \ref{table:lab_main} is not significantly different from that shown in Column (1) of Table \ref{table:lab_main}  ($p=0.90$, two-tailed Z-test), suggesting no significant influence of (superior) prior belief in the banker-AI's prediction accuracy to explain participants' greater reliance on human-AI collaborative than AI advice. Additionally, an analysis of the alternative dependent variable $GapFinalRiskAssess$ shows consistent results with the field experiment shown in Table \ref{table:field_main_robustness} Column (5). We further replicate the results of our field experiment by examining the impact of advised risk assessment (proxy to uncertainty) on participants' investment alignment advice in online experiments under the Human-AI and AI-only conditions. Our analysis indicates that higher alignment to human-AI collaborative advice primarily occurs with riskier investments, as detailed in Table \ref{table:lab_risky}, reflecting similar trends observed in Table \ref{table:field_moderation}. The detailed replication results are in Online Appendix \ref{replication}.

Taken together, the online experiment replicates the main findings of the field experiment while eliminating potential confounding factors, which we could not fully control for or eliminate in a field setting.

\subsubsection{Analysis of Potential Underlying Mechanisms}\label{mediation_results}

Evidence from field and online experiments suggests that peripheral persuasion enhances individuals' reliance on human-AI collaborative advice, particularly in situations of greater uncertainty, as proxied by investment riskiness. We further explore direct evidence of the underlying mechanism in the online experiment. In particular, we study how uncertainty, mediated by peripheral route elements such as emotional trust, relinquishment of decision-making autonomy, and shared decision responsibility, influences this increased reliance on human involvement in AI-based advice. Given that our treatment and mediators operate at the individual level (level 2), while the dependent variable is measured at the individual-loan level (level 1), we employ the multilevel moderated mediation estimation method as outlined by \cite{preacher2016multilevel}. Because the moderated mediation we seek to investigate occurs at the individual level, our moderator must also be at this level. To proxy a participant's general decision uncertainty, we calculate the percentage of decisions where confidence does not exceed the median evaluation across ten \textit{initial} investments before receiving advice. This measure is denoted as $PercUncertainDecision$. This more direct measure of uncertainty, instead of its proxy using investment riskiness, also affects individuals' alignment with human-involved AI compared to pure AI advice. The results are shown in Online Appendix Table \ref{table:lab_uncertain}, indicating that higher alignment with human-AI collaborative advice primarily occurs among individuals experiencing greater uncertainty.

We present the conceptual model of the potential underlying mechanism in Figure \ref{fig:lab_conceptual_model}. We first investigate whether uncertainty moderates how human involvement in AI-based advice influences persuasive elements. Next, we explore the mediating role of peripheral persuasive elements before testing the complete moderated mediation process.

\paragraph{The Moderating Role of Uncertainty on Persuasive Elements}

Given the conceptual model presented in Figure \ref{fig:lab_conceptual_model}, we first examine which persuasive elements are likely driving the alignment of final investment decisions with advice under heightened uncertainty without excluding the possibility of a central route influence. Specifically, if a persuasive element mediates how human involvement in AI advice influences individuals' willingness to follow that advice, and the level of uncertainty experienced moderates this influence, we expect that uncertainty will also moderate how human involvement in AI advice affects the persuasive element(s). The persuasive elements under consideration include cognitive trust, emotional trust, disappointment in the advisor's prediction accuracy, the tendency to relinquish decision autonomy, and perceived shared responsibility with the advisor. To assess this, we employ a regression model where each persuasive element is regressed on the treatment variable, moderated by $PercUncertainDecision$ (mean-centered) while controlling for individual-level variables with random effects at the individual level. The OLS regression results in Table \ref{table:lab_med_a} reveal that decision uncertainty significantly moderates perceived emotional trust. We note that decision uncertainty does not moderate central route persuasion elements, i.e., cognitive trust and disappointment in the advisor's prediction accuracy. This finding suggests that human involvement engenders greater emotional trust in the AI-based solution via the peripheral route, particularly in scenarios of heightened decision uncertainty, as individuals could derive greater decision comfort by relying on the bank advisor \citep{loewenstein2001risk}.

\begin{table}[h!]
    \centering
    \caption{\label{table:lab_med_a}Participants are likely to increase their emotional trust in the human advisor using AI when experiencing greater decision uncertainty -- OLS regression results}
    \begin{threeparttable}
    \scalebox{0.7}{{
\def\sym#1{\ifmmode^{#1}\else\(^{#1}\)\fi}
\begin{tabular}{l*{5}{c}}
\hline\hline
                                 &        DV: CogTrust    &        DV: EmoTrust    &  DV: Disappointment    &        DV: Autonomy    &     DV: AdvisorResp    \\
                                 \cline{2-6}
                                 &\multicolumn{1}{c}{(1)}    &\multicolumn{1}{c}{(2)}    &\multicolumn{1}{c}{(3)}    &\multicolumn{1}{c}{(4)}    &\multicolumn{1}{c}{(5)}    \\
\hline
Human-AI (vs. AI-only)                &               0.339    &               0.305    &               0.164    &              -0.185    &               0.136    \\
                                 &             (0.200)    &             (0.246)    &             (0.399)    &             (0.403)    &             (0.092)    \\
PercUncertainDecision            &              -0.706    &              -1.844    &              -1.409    &              -1.074    &              -0.217    \\
                                 &             (0.780)    &             (0.955)    &             (1.552)    &             (1.567)    &             (0.357)    \\
Human-AI (vs. AI-only)$\times$PercUncertainDecision&               1.192    &               3.554**  &              -0.355    &               3.955    &               0.490    \\
                                 &             (1.130)    &             (1.385)    &             (2.250)    &             (2.272)    &             (0.518)    \\
Constant                         &               4.658****&               3.485*** &               0.891    &               1.136    &              -0.331    \\
                                 &             (1.022)    &             (1.252)    &             (2.035)    &             (2.055)    &             (0.468)    \\
\hline
Observations                     &                  87    &                  87    &                  87    &                  87    &                  87    \\
\(R^{2}\)                        &               0.268    &               0.248    &               0.248    &               0.252    &               0.192    \\
Date fixed-effects               &                 Yes    &                 Yes    &                 Yes    &                 Yes    &                 Yes    \\
Ind-level ctrls                  &                 Yes    &                 Yes    &                 Yes    &                 Yes    &                 Yes    \\
\hline\hline
\end{tabular}
}
}
    \begin{tablenotes}[flushleft]
    \footnotesize
        \item\leavevmode\kern-\scriptspace\kern-\labelsep\textit{Note}: Standard errors clustered at the individual level are shown in the parentheses; **$p<0.05$, ***$p<0.01$, \\ ****$p<0.001$.
                
    \end{tablenotes}
    \end{threeparttable}
\end{table}

\paragraph{The Mediating Role of Peripheral Persuasive Elements on Advice-taking}
Following the results shown in Table \ref{table:lab_med_a}, we test the potential mediating role of emotional trust by incorporating participants' emotional trust into the model specification outlined in Table \ref{table:lab_main} Column (1). The OLS regression results, presented in Column (2) of Table \ref{table:lab_med_all}, reveal that emotional trust significantly contributes to greater alignment of final investment decisions with the advice provided. For comparison, Column (1) of Table \ref{table:lab_med_all} displays the results from Table \ref{table:lab_main}, where we observe a reduction in the magnitude of the main treatment effect coefficient ($\beta_1$).

\begin{table}[h!]
    \centering
    \caption{\label{table:lab_med_all}Participants' emotional trust positively correlates with their alignment in the final investments with advice -- OLS regression results}
    \begin{threeparttable}
    \footnotesize
    {
\fontsize{8}{9}\selectfont
\def\sym#1{\ifmmode^{#1}\else\(^{#1}\)\fi}
\begin{tabular}{l*{2}{c}}
\hline\hline
                                 &     \multicolumn{2}{c}{DV: FinalAlign}                         \\
                                 \cline{2-3}
                                 &\multicolumn{1}{c}{(1)}    &\multicolumn{1}{c}{(2)}    \\
\hline
Human-AI (vs. AI)                &               0.071**  &               0.059**  \\
                                 &             (0.030)    &             (0.026)    \\
EmoTrust                         &                        &               0.041****\\
                                 &                        &             (0.011)    \\
Constant                         &               0.661****&               0.494****\\
                                 &             (0.021)    &             (0.047)    \\
\hline
Observations                     &                 870    &                 870    \\
\(R^{2}\)                        &               0.422    &               0.429    \\
Loan fixed-effects               &                 Yes    &                 Yes    \\
Date fixed-effects               &                 Yes    &                 Yes    \\
Controls                         &                 Yes    &                 Yes    \\
\hline\hline
\end{tabular}
}

    \begin{tablenotes}[flushleft]
    \footnotesize
        \item\leavevmode\kern-\scriptspace\kern-\labelsep\textit{Note}: Standard errors clustered at the individual level are shown in the parentheses; **$p<0.05$, ***$p<0.01$, ****$p<0.001$.
        
    \end{tablenotes}
    \end{threeparttable}
\end{table}

\paragraph{Moderated Mediation Results}
Next, we formally test the indirect effect of emotional trust on the alignment of final investments with advice moderated by uncertainty. High and low uncertainty levels are defined as one standard deviation above and below the mean, respectively, following \citep{hayes2017introduction}. We estimate the multilevel (2-2-1) moderated mediation model \citep[Model 7;][]{hayes2017introduction} using the \texttt{lavaan} package in R that accounts for the random effects on the individual level.\footnote{Our estimated model is conceptually similar to Model 7 introduced by \cite{hayes2017introduction}, except for involving multi-level estimation, which the PROCESS Macro does not account for.} Our findings indicate that involving a human banker in AI-based advice, compared to AI-only advice, has an indirect effect of increasing alignment with the advice under conditions of higher decision uncertainty, mediated by emotional trust (effect size = $0.040$, 95\% confidence interval = $[0.005, 0.076]$, $p<0.05$). This positive mediation effect is not observed when participants are more certain about their decisions (effect size = $-0.015$, 95\% confidence interval = $[-0.042, 0.012]$, $p=0.27$). Furthermore, we identify a significant moderated mediation effect, indicating that the positive mediation effect of emotional trust under greater uncertainty is significantly stronger than under lower uncertainty (effect size = $0.056$, 95\% confidence interval = $[0.006, 0.105]$, $p<0.05$).

\subsection{Discussion of online controlled experiment}

Our controlled and incentivized online experiment serves three purposes. First, we confirm the main findings of the field experiment: individuals are more likely to follow investment recommendations when human bankers are involved in the AI-based solution. The replication of the main results reduces concerns that these results could be influenced by factors such as the advisor's description, the individually assumed prediction accuracy, or the information inputs used by the banker and AI in generating advice. %---all of which we can control for in the complementary experiment. 
Second, we assess participants' prior beliefs about the prediction accuracy of the banker, AI, and banker-AI combination, alongside their perceived disappointment in the advisor's accuracy after evaluations. The evidence indicates that neither prior beliefs nor disappointment in prediction accuracy can adequately explain the positive response to the involvement of human bankers in AI-based investment recommendations. We also did not find supporting evidence that the relatively low reliance on pure AI advice is due to a perceived poor performance of the AI; participants' belief in AI's prediction accuracy is not different from 70\% as communicated in the field experiment ($p=0.54$ using a one-sample Student's t-test). Interestingly, participants believe banker's prediction accuracy is significantly lower than 70\% ($p<0.01$ using a one-sample Student's t-test), suggesting the greater alignment of the final investment with human-only advice in the field experiment is \textit{unlikely} driven by a greater prior belief in the prediction performance from a banker than the communicated AI performance at 70\%. Third, through moderated mediation analysis, we uncover that participants' positive response to the inclusion of a banker in AI-based solutions is primarily driven by increased emotional trust in the banker via the peripheral route, especially when experiencing heightened investment uncertainty. These findings suggest that incorporating human elements into AI-based decision-making processes can enhance consumer comfort in uncertain situations. However, while high-quality advice benefits consumers, poor-quality service may mislead them.

\section{Discussion and Conclusion}

With rapid advancements in AI technologies, the automation of service provision is becoming increasingly viable across industries, promising efficiency gains through AI's superior task performance and lower operating costs \citep{agrawal2022prediction}. However, fully AI-driven services are not without risks. Concerns surrounding accountability, ethics, and legality have sparked ongoing debates among policymakers, practitioners, and scholars regarding the necessity of human involvement in AI-based services \citep{buckley2021regulating, europeancommission2019}. While much of this discourse has centered on how human-AI collaboration affects service quality, comparatively less attention has been given to how consumers respond to human involvement in AI-driven services. Our study addresses this gap. In a field study conducted in collaboration with an industry partner, we find that bank customers exhibit greater reliance on investment advice when they are aware of human involvement in the advisory process, an effect that, in our setting, enhances consumer welfare. This increased reliance is not driven by perceptions of superior advice quality due to human-AI complementarity. %AI-assisted human expertise. 
Instead, the results from our field experiment and a complementary online experiment consistently indicate that human involvement shapes consumer behavior through the peripheral route of persuasion. 

Our findings have important implications for companies contemplating to offer fully-automated or collaborative human-AI services. First, while maintaining human oversight may arguably incur additional costs, our findings suggest that human involvement increases the persuasiveness of AI-generated recommendations. Consumers exhibit greater reliance on human-AI collaborative advice compared to AI-only recommendations which can yield better consumer outcomes if the resulting advice is more accurate. Over time, stronger engagement and improved decision-making may lead to a higher willingness to pay, potentially justifying the costs associated with maintaining human oversight. Second, our results indicate that consumers' positive response to human involvement in AI-based services is not primarily driven by their perception of enhanced advice quality. It seems that an affective reassurance provided by human presence as such can drive an increase in advice utilization through the peripheral route of persuasion. This highlights that humans can complement AI-driven services even when they do not enhance technical quality---a subtle yet pivotal consideration for firms evaluating the financial viability of collaborative service models. Last but not least, our findings caution that keeping humans in the loop may also introduce risks for consumers if the resulting human-AI collaboration leads to suboptimal service quality. As such, a human in the loop may foster over-reliance when trust is misplaced. This risk underscores the need for ongoing quality monitoring in human-AI collaborative services. Firms must ensure that human involvement does not inadvertently lead consumers to rely more frequently on poor-quality outcomes, particularly in high-stakes domains where incorrect recommendations carry notable consequences.

Our study has limitations that we believe provide avenues for future research. First, our investigation focuses on the investment advisory domain. While we are confident that our findings regarding the peripheral mechanism represent a fundamental psychological mechanism applicable to different decision domains, additional contextual factors, such as task importance, may shape consumer responses. In high-stakes financial, medical, or legal decisions, human oversight may offer stronger emotional reassurance than in lower-stakes contexts. Identifying these contextual nuances and examining how they shape consumer preferences for human-AI collaborative versus purely AI-driven solutions represents a promising avenue for future research.

Second, our study compares human-AI collaboration to a pure AI service without a genuine ``human-only" condition. In particular, our paper's ``human-only" condition conceals AI's contribution to human-AI collaborative advice, meaning we cannot conclude whether human-AI collaboration is superior to purely human advice. Advisory services offered exclusively by human professionals may continue to represent the prevailing standard in certain areas of decision-making, such as wealth management and ethical investment, probably due to consumers' trust in human expertise alone. Therefore, a promising direction for future research is to explore how human-AI collaboration fares against human-only services regarding quality and consumer reception.

Third, our empirical studies feature an AI and a collaborative human-AI system with approximately 70\% accuracy, effectively representing a mid-level prediction quality in the context of financial risk scoring \citep{lessmann2015benchmarking, dong2024accuracy}. Previous literature suggests that higher AI prediction accuracy could engender greater trust in the AI service \citep{yin2019understanding}, thereby positively impacting individuals' reception of the pure AI advice \citep[e.g.,][]{you2022algorithmic}. Against this background, it seems possible that greater human-AI collaborative prediction accuracy would further increase consumers' positive response to the human-AI collaborative service. Moreover, AI accuracy may influence the recorded treatment effects, as the extent of human persuasion could become less relevant when the AI’s predictive accuracy is higher, and more relevant when it is lower performance. Such a moderation is particularly relevant when AI capabilities continue to grow. We leave such an investigation to future research.

Finally, given findings in prior research suggesting that human advisor's exhibited confidence increases advice-taking \citep[e.g.,][]{sniezek1995cueing,sniezek2001trust}, it remains an open question whether a more confident human presence strengthens the persuasive effect of human-AI collaboration.  Related to the role of human advisor confidence, the lower reliance on pure AI advice may diminish if the AI reports predictions together with confidence scores \citep[e.g.,][]{taudien2022calibrating}, signaling to the consumer how ``certain the AI is." While our paper does not focus on these aspects, we believe they offer fruitful avenues for future research on the consumption-side effects of human-AI collaboration.

Notwithstanding certain limitations, our research demonstrates that the synergistic nature of human-AI collaboration transcends mere improvements in task performance, extending to encompass psychological and behavioral factors that significantly influence consumer decisions. We anticipate that this fundamental finding will incite further exploration into the psychological implications of human participation, thus aiding firms and policymakers in making more informed decisions regarding AI-enabled services.

\newpage
\singlespacing
\bibliographystyle{apalike}
\bibliography{references}

\begin{thebibliography}{}

\bibitem[Abdel-Karim et~al., 2023]{abdel2023ai}
Abdel-Karim, B.~M., Pfeuffer, N., Carl, K.~V., and Hinz, O. (2023).
\newblock How ai-based systems can induce reflections: The case of ai-augmented
  diagnostic work.
\newblock {\em MIS Quarterly}, 47(4):1395--1424.

\bibitem[Acemoglu and Restrepo, 2019]{acemoglu2019automation}
Acemoglu, D. and Restrepo, P. (2019).
\newblock Automation and new tasks: How technology displaces and reinstates
  labor.
\newblock {\em Journal of Economic Perspectives}, 33(2):3--30.

\bibitem[Agrawal et~al., 2024]{agrawal2022prediction}
Agrawal, A., Gans, J.~S., and Goldfarb, A. (2024).
\newblock Prediction machines, insurance, and protection: An alternative
  perspective on ai’s role in production.
\newblock {\em Journal of the Japanese and International Economies}, 72:101307.

\bibitem[Angrist and Pischke, 2009]{angrist2009mostly}
Angrist, J.~D. and Pischke, J.-S. (2009).
\newblock {\em Mostly harmless econometrics: An empiricist's companion}.
\newblock Princeton university press.

\bibitem[Asch, 1955]{asch1955opinions}
Asch, S.~E. (1955).
\newblock Opinions and social pressure.
\newblock {\em Scientific American}, 193(5):31--35.

\bibitem[Avramov et~al., 2023]{avramov2022machine}
Avramov, D., Cheng, S., and Metzker, L. (2023).
\newblock Machine learning vs. economic restrictions: Evidence from stock
  return predictability.
\newblock {\em Management Science}, 69(5):2587--2619.

\bibitem[Awad et~al., 2018]{awad2018moral}
Awad, E., Dsouza, S., Kim, R., Schulz, J., Henrich, J., Shariff, A., Bonnefon,
  J.-F., and Rahwan, I. (2018).
\newblock The moral machine experiment.
\newblock {\em Nature}, 563(7729):59--64.

\bibitem[Bauer et~al., 2024a]{bauer2024feedback}
Bauer, K., Heigl, R., Hinz, O., and Kosfeld, M. (2024a).
\newblock Feedback loops in machine learning: A study on the interplay of
  continuous updating and human discrimination.
\newblock {\em Journal of the Association for Information Systems},
  25(4):804--866.

\bibitem[Bauer et~al., 2024b]{bauer2024all}
Bauer, K., Jussupow, E., Heigl, R., Vogt, B., and Hinz, O. (2024b).
\newblock All just in your head? unraveling the side effects of generative ai
  disclosure in creative task (february 29, 2024).
\newblock Working paper, SSRN.
\newblock Available at SSRN:
  \url{https://papers.ssrn.com/sol3/papers.cfm?abstract_id=4782554}.

\bibitem[Bauer et~al., 2023]{bauer2023expl}
Bauer, K., von Zahn, M., and Hinz, O. (2023).
\newblock Expl (ai) ned: The impact of explainable artificial intelligence on
  users’ information processing.
\newblock {\em Information systems research}, 34(4):1582--1602.

\bibitem[Bergstein, 2020]{bergstein2020ai}
Bergstein, B. (2020).
\newblock What ai still can’t do.
\newblock {\em MIT Technology Review}, 123(2):1--7.

\bibitem[Berk, 2017]{berk2017impact}
Berk, R. (2017).
\newblock An impact assessment of machine learning risk forecasts on parole
  board decisions and recidivism.
\newblock {\em Journal of Experimental Criminology}, 13(2):193--216.

\bibitem[Bonaccio and Dalal, 2006]{bonaccio2006advice}
Bonaccio, S. and Dalal, R.~S. (2006).
\newblock Advice taking and decision-making: An integrative literature review,
  and implications for the organizational sciences.
\newblock {\em Organizational Behavior and Human Decision Processes},
  101(2):127--151.

\bibitem[Brynjolfsson et~al., 2018]{brynjolfsson2018can}
Brynjolfsson, E., Mitchell, T., and Rock, D. (2018).
\newblock What can machines learn, and what does it mean for occupations and
  the economy?
\newblock {\em AEA Papers and Proceedings}, 108:43--47.

\bibitem[Buckley et~al., 2021]{buckley2021regulating}
Buckley, R.~P., Zetzsche, D.~A., Arner, D.~W., and Tang, B.~W. (2021).
\newblock Regulating artificial intelligence in finance: Putting the human in
  the loop.
\newblock {\em The Sydney Law Review}, 43(1):43--81.

\bibitem[Cao et~al., 2024]{cao2021man}
Cao, S., Jiang, W., Wang, J., and Yang, B. (2024).
\newblock From man vs. machine to man+ machine: The art and ai of stock
  analyses.
\newblock {\em Journal of Financial Economics}, 160:103910.

\bibitem[Castelo et~al., 2019]{castelo2019task}
Castelo, N., Bos, M.~W., and Lehmann, D.~R. (2019).
\newblock Task-dependent algorithm aversion.
\newblock {\em Journal of Marketing Research}, 56(5):809--825.

\bibitem[Chang et~al., 2022]{chang2022machine}
Chang, A.-H., Yang, L.-K., Tsaih, R.-H., and Lin, S.-K. (2022).
\newblock Machine learning and artificial neural networks to construct p2p
  lending credit-scoring model: A case using lending club data.
\newblock {\em Quantitative Finance and Economics}, 6(2):303--325.

\bibitem[Commission, 2019]{europeancommission2019}
Commission, E. (2019).
\newblock Communication from the commission to the european parliament, the
  council, the european economic and social committee and the committee of the
  regions - building trust in human centric artificial intelligence
  (com(2019)168).
\newblock
  \url{https://digital-strategy.ec.europa.eu/en/library/communication-building-trust-human-centric-artificial-intelligence}.
\newblock Accessed: 2024-08-29.

\bibitem[Cooper et~al., 1988]{cooper1988entrepreneurs}
Cooper, A.~C., Woo, C.~Y., and Dunkelberg, W.~C. (1988).
\newblock Entrepreneurs' perceived chances for success.
\newblock {\em Journal of Business Venturing}, 3(2):97--108.

\bibitem[Costa and Henshaw, 2022]{costa2022quantifying}
Costa, P. and Henshaw, J.~E. (2022).
\newblock Quantifying the investor’s view on the value of human and
  robo-advice.
\newblock {\em Vanguard Research Report}, (1).

\bibitem[Cowgill and Tucker, 2020]{cowgill2020algorithmic}
Cowgill, B. and Tucker, C.~E. (2020).
\newblock Algorithmic fairness and economics.
\newblock Working paper, Columbia Business School Research Paper.
\newblock Available at SSRN: \url{https://ssrn.com/abstract=3361280} or
  \url{http://dx.doi.org/10.2139/ssrn.3361280}.

\bibitem[Curran, 2016]{curran2016methods}
Curran, P.~G. (2016).
\newblock Methods for the detection of carelessly invalid responses in survey
  data.
\newblock {\em Journal of Experimental Social Psychology}, 66:4--19.

\bibitem[Dalal and Bonaccio, 2010]{dalal2010types}
Dalal, R.~S. and Bonaccio, S. (2010).
\newblock What types of advice do decision-makers prefer?
\newblock {\em Organizational Behavior and Human Decision Processes},
  112(1):11--23.

\bibitem[Darda and Cross, 2023]{darda2023computer}
Darda, K.~M. and Cross, E.~S. (2023).
\newblock The computer, a choreographer? aesthetic responses to
  randomly-generated dance choreography by a computer.
\newblock {\em Heliyon}, 9(1).

\bibitem[Dell’Acqua, 2022]{dell2021falling}
Dell’Acqua, F. (2022).
\newblock Falling asleep at the wheel: Human/ai collaboration in a field
  experiment on hr recruiters.
\newblock Technical report, Working paper.

\bibitem[Dietvorst et~al., 2015]{dietvorst2015algorithm}
Dietvorst, B.~J., Simmons, J.~P., and Massey, C. (2015).
\newblock Algorithm aversion: People erroneously avoid algorithms after seeing
  them err.
\newblock {\em Journal of Experimental Psychology: General}, 144(1):114.

\bibitem[Dohmen et~al., 2012]{dohmen2012intergenerational}
Dohmen, T., Falk, A., Huffman, D., and Sunde, U. (2012).
\newblock The intergenerational transmission of risk and trust attitudes.
\newblock {\em The Review of Economic Studies}, 79(2):645--677.

\bibitem[Dong et~al., 2024]{dong2024accuracy}
Dong, H., Liu, R., and Tham, A.~W. (2024).
\newblock Accuracy comparison between five machine learning algorithms for
  financial risk evaluation.
\newblock {\em Journal of Risk and Financial Management}, 17(2):50.

\bibitem[EU, 2024]{EU2024}
EU (2024).
\newblock Proposal for a regulation of the european parliament and of the
  council laying down harmonised rules on artificial intelligence (artificial
  intelligence act) and amending certain union legislative acts.
\newblock
  \url{https://data.consilium.europa.eu/doc/document/ST-5662-2024-INIT/en/pdf}.
\newblock Accessed: 2024-08-12.

\bibitem[Faisal et~al., 2021]{faisal2021credit}
Faisal, M.~F., Saqlain, M. N.~U., Bhuiyan, M. A.~S., Miraz, M.~H., and Patwary,
  M.~J. (2021).
\newblock Credit approval system using machine learning: Challenges and future
  directions.
\newblock In {\em 2021 International Conference on Computing, Networking,
  Telecommunications \& Engineering Sciences Applications (CoNTESA)}, pages
  76--82. IEEE.

\bibitem[Feuerriegel et~al., 2022]{feuerriegel2022bringing}
Feuerriegel, S., Shrestha, Y.~R., von Krogh, G., and Zhang, C. (2022).
\newblock Bringing artificial intelligence to business management.
\newblock {\em Nature Machine Intelligence}, 4(7):611--613.

\bibitem[Friestad and Wright, 1994]{friestad1994persuasion}
Friestad, M. and Wright, P. (1994).
\newblock The persuasion knowledge model: How people cope with persuasion
  attempts.
\newblock {\em Journal of Consumer Research}, 21(1):1--31.

\bibitem[F{\"u}gener et~al., 2021]{fugener2021will}
F{\"u}gener, A., Grahl, J., Gupta, A., and Ketter, W. (2021).
\newblock Will humans-in-the-loop become borgs? merits and pitfalls of working
  with ai.
\newblock {\em MIS Quarterly}, 45(3):1527--1556.

\bibitem[F{\"u}gener et~al., 2022]{fugener2021cognitive}
F{\"u}gener, A., Grahl, J., Gupta, A., and Ketter, W. (2022).
\newblock Cognitive challenges in human--artificial intelligence collaboration:
  Investigating the path toward productive delegation.
\newblock {\em Information Systems Research}, 33(2):678--696.

\bibitem[Goh et~al., 2024]{goh2024large}
Goh, E., Gallo, R., Hom, J., Strong, E., Weng, Y., Kerman, H., Cool, J.~A.,
  Kanjee, Z., Parsons, A.~S., Ahuja, N., et~al. (2024).
\newblock Large language model influence on diagnostic reasoning: a randomized
  clinical trial.
\newblock {\em JAMA Network Open}, 7(10):e2440969--e2440969.

\bibitem[Greene, 2004]{greene2004fixed}
Greene, W. (2004).
\newblock Fixed effects and bias due to the incidental parameters problem in
  the tobit model.
\newblock {\em Econometric reviews}, 23(2):125--147.

\bibitem[Harvey and Fischer, 1997]{harvey1997taking}
Harvey, N. and Fischer, I. (1997).
\newblock Taking advice: Accepting help, improving judgment, and sharing
  responsibility.
\newblock {\em Organizational Behavior and Human Decision Processes},
  70(2):117--133.

\bibitem[Hayes, 2017]{hayes2017introduction}
Hayes, A.~F. (2017).
\newblock {\em Introduction to mediation, moderation, and conditional process
  analysis: A regression-based approach}.
\newblock Guilford publications.

\bibitem[Hermosilla et~al., 2018]{hermosilla2018can}
Hermosilla, M., Guti{\'e}rrez-Navratil, F., and Prieto-Rodriguez, J. (2018).
\newblock Can emerging markets tilt global product design? impacts of chinese
  colorism on hollywood castings.
\newblock {\em Marketing Science}, 37(3):356--381.

\bibitem[Kahneman et~al., 2016]{kahneman2016noise}
Kahneman, D., Rosenfield, A.~M., Gandhi, L., and Blaser, T. (2016).
\newblock Noise: How to overcome the high, hidden cost of inconsistent decision
  making.
\newblock {\em Harvard business review}, 94(10):38--46.

\bibitem[Kellogg et~al., 2020]{kellogg2020algorithms}
Kellogg, K.~C., Valentine, M.~A., and Christin, A. (2020).
\newblock Algorithms at work: The new contested terrain of control.
\newblock {\em Academy of management annals}, 14(1):366--410.

\bibitem[Kingston, 2016]{kingston2016artificial}
Kingston, J.~K. (2016).
\newblock Artificial intelligence and legal liability.
\newblock In {\em International Conference on Innovative Techniques and
  Applications of Artificial Intelligence XXIV}, number~33, pages 269--279.
  Springer.

\bibitem[Koestner et~al., 1999]{koestner1999follow}
Koestner, R., Gingras, I., Abutaa, R., Losier, G.~F., DiDio, L., and Gagn{\'e},
  M. (1999).
\newblock To follow expert advice when making a decision: An examination of
  reactive versus reflective autonomy.
\newblock {\em Journal of Personality}, 67(5):851--872.

\bibitem[Komiak and Benbasat, 2006]{komiak2006effects}
Komiak, S.~Y. and Benbasat, I. (2006).
\newblock The effects of personalization and familiarity on trust and adoption
  of recommendation agents.
\newblock {\em MIS Quarterly}, 30(4):941--960.

\bibitem[Kumar et~al., 2016]{kumar2016credit}
Kumar, V., Natarajan, S., Keerthana, S., Chinmayi, K., and Lakshmi, N. (2016).
\newblock Credit risk analysis in peer-to-peer lending system.
\newblock In {\em 2016 IEEE international conference on knowledge engineering
  and applications (ICKEA)}, pages 193--196. IEEE.

\bibitem[Lebovitz et~al., 2022]{lebovitz2022engage}
Lebovitz, S., Lifshitz-Assaf, H., and Levina, N. (2022).
\newblock To engage or not to engage with ai for critical judgments: How
  professionals deal with opacity when using ai for medical diagnosis.
\newblock {\em Organization Science}, 33(1):126--148.

\bibitem[Lessmann et~al., 2015]{lessmann2015benchmarking}
Lessmann, S., Baesens, B., Seow, H.-V., and Thomas, L.~C. (2015).
\newblock Benchmarking state-of-the-art classification algorithms for credit
  scoring: An update of research.
\newblock {\em European Journal of Operational Research}, 247(1):124--136.

\bibitem[Li et~al., 2022]{li2022recommender}
Li, X., Grahl, J., and Hinz, O. (2022).
\newblock How do recommender systems lead to consumer purchases? a causal
  mediation analysis of a field experiment.
\newblock {\em Information Systems Research}, 33(2):620--637.

\bibitem[Loewenstein et~al., 2001]{loewenstein2001risk}
Loewenstein, G.~F., Weber, E.~U., Hsee, C.~K., and Welch, N. (2001).
\newblock Risk as feelings.
\newblock {\em Psychological Bulletin}, 127(2):267.

\bibitem[Longoni et~al., 2019]{longoni2019resistance}
Longoni, C., Bonezzi, A., and Morewedge, C.~K. (2019).
\newblock Resistance to medical artificial intelligence.
\newblock {\em Journal of Consumer Research}, 46(4):629--650.

\bibitem[Lu and Zhang, 2024]{lu20241+}
Lu, T. and Zhang, Y. (2024).
\newblock 1+1> 2? information, humans, and machines.
\newblock {\em Information Systems Research}, forthcoming.

\bibitem[MacCrimmon and Wehrung, 1990]{maccrimmon1990characteristics}
MacCrimmon, K.~R. and Wehrung, D.~A. (1990).
\newblock Characteristics of risk taking executives.
\newblock {\em Management science}, 36(4):422--435.

\bibitem[Malekipirbazari and Aksakalli, 2015]{malekipirbazari2015risk}
Malekipirbazari, M. and Aksakalli, V. (2015).
\newblock Risk assessment in social lending via random forests.
\newblock {\em Expert Systems with Applications}, 42(10):4621--4631.

\bibitem[March and Shapira, 1987]{march1987managerial}
March, J.~G. and Shapira, Z. (1987).
\newblock Managerial perspectives on risk and risk taking.
\newblock {\em Management Science}, 33(11):1404--1418.

\bibitem[MarketsandMarkets, 2025]{marketsandmarketsFinanceMarket}
MarketsandMarkets (2025).
\newblock {A}{I} in {F}inance {M}arket {S}ize, {S}hare, {G}rowth {R}eport -
  2030 --- marketsandmarkets.com.
\newblock
  \url{https://www.marketsandmarkets.com/Market-Reports/ai-in-finance-market-90552286.html}.
\newblock [Accessed 03-03-2025].

\bibitem[McKnight et~al., 2002]{mcknight2002developing}
McKnight, D.~H., Choudhury, V., and Kacmar, C. (2002).
\newblock Developing and validating trust measures for e-commerce: An
  integrative typology.
\newblock {\em Information Systems Research}, 13(3):334--359.

\bibitem[Meade and Craig, 2012]{meade2012identifying}
Meade, A.~W. and Craig, S.~B. (2012).
\newblock Identifying careless responses in survey data.
\newblock {\em Psychological Methods}, 17(3):437.

\bibitem[Millet et~al., 2023]{millet2023defending}
Millet, K., Buehler, F., Du, G., and Kokkoris, M.~D. (2023).
\newblock Defending humankind: Anthropocentric bias in the appreciation of ai
  art.
\newblock {\em Computers in Human Behavior}, 143:107707.

\bibitem[Mishra and Metilda, 2015]{mishra2015study}
Mishra, K. and Metilda, M.~J. (2015).
\newblock A study on the impact of investment experience, gender, and level of
  education on overconfidence and self-attribution bias.
\newblock {\em IIMB Management Review}, 27(4):228--239.

\bibitem[{\"O}nkal et~al., 2009]{onkal2009relative}
{\"O}nkal, D., Goodwin, P., Thomson, M., G{\"o}n{\"u}l, S., and Pollock, A.
  (2009).
\newblock The relative influence of advice from human experts and statistical
  methods on forecast adjustments.
\newblock {\em Journal of Behavioral Decision Making}, 22(4):390--409.

\bibitem[Petty and Cacioppo, 1986]{petty1986elaboration}
Petty, R.~E. and Cacioppo, J.~T. (1986).
\newblock The elaboration likelihood model of persuasion.
\newblock In {\em Communication and Persuasion}, pages 1--24. Springer.

\bibitem[Pfeuffer et~al., 2023]{pfeuffer2023explanatory}
Pfeuffer, N., Baum, L., Stammer, W., Abdel-Karim, B.~M., Schramowski, P.,
  Bucher, A.~M., H{\"u}gel, C., Rohde, G., Kersting, K., and Hinz, O. (2023).
\newblock Explanatory interactive machine learning: Establishing an action
  design research process for machine learning projects.
\newblock {\em Business \& Information Systems Engineering}, 65(6):677--701.

\bibitem[Preacher et~al., 2016]{preacher2016multilevel}
Preacher, K.~J., Zhang, Z., and Zyphur, M.~J. (2016).
\newblock Multilevel structural equation models for assessing moderation within
  and across levels of analysis.
\newblock {\em Psychological Methods}, 21(2):189.

\bibitem[Rahwan et~al., 2019]{rahwan2019machine}
Rahwan, I., Cebrian, M., Obradovich, N., Bongard, J., Bonnefon, J.-F.,
  Breazeal, C., Crandall, J.~W., Christakis, N.~A., Couzin, I.~D., Jackson,
  M.~O., et~al. (2019).
\newblock Machine behaviour.
\newblock {\em Nature}, 568(7753):477--486.

\bibitem[Rohaan et~al., 2022]{rohaan2022using}
Rohaan, D., Topan, E., and Groothuis-Oudshoorn, C.~G. (2022).
\newblock Using supervised machine learning for b2b sales forecasting: A case
  study of spare parts sales forecasting at an after-sales service provider.
\newblock {\em Expert Systems with Applications}, 188:115925.

\bibitem[Singh et~al., 2019]{singh2019automating}
Singh, R., Ayyar, M.~P., Pavan, T. V.~S., Gosain, S., and Shah, R.~R. (2019).
\newblock Automating car insurance claims using deep learning techniques.
\newblock In {\em 2019 IEEE Fifth International Conference on Multimedia Big
  Data (BigMM)}, pages 199--207. IEEE.

\bibitem[Slovic, 1987]{slovic1987perception}
Slovic, P. (1987).
\newblock Perception of risk.
\newblock {\em Science}, 236(4799):280--285.

\bibitem[Sniezek and Buckley, 1995]{sniezek1995cueing}
Sniezek, J.~A. and Buckley, T. (1995).
\newblock Cueing and cognitive conflict in judge-advisor decision making.
\newblock {\em Organizational Behavior and Human Decision Processes},
  62(2):159--174.

\bibitem[Sniezek and Van~Swol, 2001]{sniezek2001trust}
Sniezek, J.~A. and Van~Swol, L.~M. (2001).
\newblock Trust, confidence, and expertise in a judge-advisor system.
\newblock {\em Organizational Behavior and Human Decision Processes},
  84(2):288--307.

\bibitem[Steffel et~al., 2016]{steffel2016passing}
Steffel, M., Williams, E.~F., and Perrmann-Graham, J. (2016).
\newblock Passing the buck: Delegating choices to others to avoid
  responsibility and blame.
\newblock {\em Organizational Behavior and Human Decision Processes},
  135:32--44.

\bibitem[Strickland et~al., 1966]{strickland1966temporal}
Strickland, L.~H., Lewicki, R.~J., and Katz, A.~M. (1966).
\newblock Temporal orientation and perceived control as determinants of
  risk-taking.
\newblock {\em Journal of Experimental Social Psychology}, 2(2):143--151.

\bibitem[Sun et~al., 2022]{sun2022predicting}
Sun, J., Zhang, D.~J., Hu, H., and Van~Mieghem, J.~A. (2022).
\newblock Predicting human discretion to adjust algorithmic prescription: A
  large-scale field experiment in warehouse operations.
\newblock {\em Management Science}, 68(2):846--865.

\bibitem[Taudien et~al., 2022]{taudien2022calibrating}
Taudien, A., Fuegener, A., Gupta, A., and Ketter, W. (2022).
\newblock Calibrating users’ mental models for delegation to {AI}.
\newblock {\em International Conference of Information Systems (ICIS)}.

\bibitem[Tversky and Kahneman, 1974]{tversky1974judgment}
Tversky, A. and Kahneman, D. (1974).
\newblock Judgment under uncertainty: Heuristics and biases.
\newblock {\em Science}, 185(4157):1124--1131.

\bibitem[von Zahn et~al., 2024]{von2024smart}
von Zahn, M., Bauer, K., Mihale-Wilson, C., Jagow, J., Speicher, M., and Hinz,
  O. (2024).
\newblock Smart green nudging: Reducing product returns through digital
  footprints and causal machine learning.
\newblock {\em Marketing Science}, forthcoming.

\bibitem[Wasserbacher and Spindler, 2022]{wasserbacher2022machine}
Wasserbacher, H. and Spindler, M. (2022).
\newblock Machine learning for financial forecasting, planning and analysis:
  recent developments and pitfalls.
\newblock {\em Digital Finance}, 4(1):63--88.

\bibitem[Weber et~al., 2002]{weber2002domain}
Weber, E.~U., Blais, A.-R., and Betz, N.~E. (2002).
\newblock A domain-specific risk-attitude scale: Measuring risk perceptions and
  risk behaviors.
\newblock {\em Journal of Behavioral Decision Making}, 15(4):263--290.

\bibitem[Weber and Milliman, 1997]{weber1997perceived}
Weber, E.~U. and Milliman, R.~A. (1997).
\newblock Perceived risk attitudes: Relating risk perception to risky choice.
\newblock {\em Management Science}, 43(2):123--144.

\bibitem[Wooldridge, 2010]{wooldridge2010econometric}
Wooldridge, J.~M. (2010).
\newblock {\em Econometric analysis of cross section and panel data}.
\newblock MIT press.

\bibitem[Yaniv, 2004]{yaniv2004receiving}
Yaniv, I. (2004).
\newblock Receiving other people’s advice: Influence and benefit.
\newblock {\em Organizational Behavior and Human Decision Processes},
  93(1):1--13.

\bibitem[Yin et~al., 2019]{yin2019understanding}
Yin, M., Wortman~Vaughan, J., and Wallach, H. (2019).
\newblock Understanding the effect of accuracy on trust in machine learning
  models.
\newblock In {\em Proceedings of the 2019 chi conference on human factors in
  computing systems}, pages 1--12.

\bibitem[You et~al., 2022]{you2022algorithmic}
You, S., Yang, C.~L., and Li, X. (2022).
\newblock Algorithmic versus human advice: Does presenting prediction
  performance matter for algorithm appreciation?
\newblock {\em Journal of Management Information Systems}, 39(2):336--365.

\end{thebibliography}

\newpage

\clearpage
\section*{Online Appendix}

\renewcommand{\thesubsection}{\Alph{subsection}}
\subsection{Additional Figures and Tables}
\setcounter{table}{0}
\setcounter{figure}{0}
\renewcommand{\thetable}{A\arabic{table}}
\renewcommand{\thefigure}{A\arabic{figure}}

\begin{figure}[htpb]
    \centering
    \caption{\label{fig:stage1_interface}The human-AI collaboration interface -- bank managers make investment recommendations and risk assessments with AI output.}
    \includegraphics[scale=0.5]{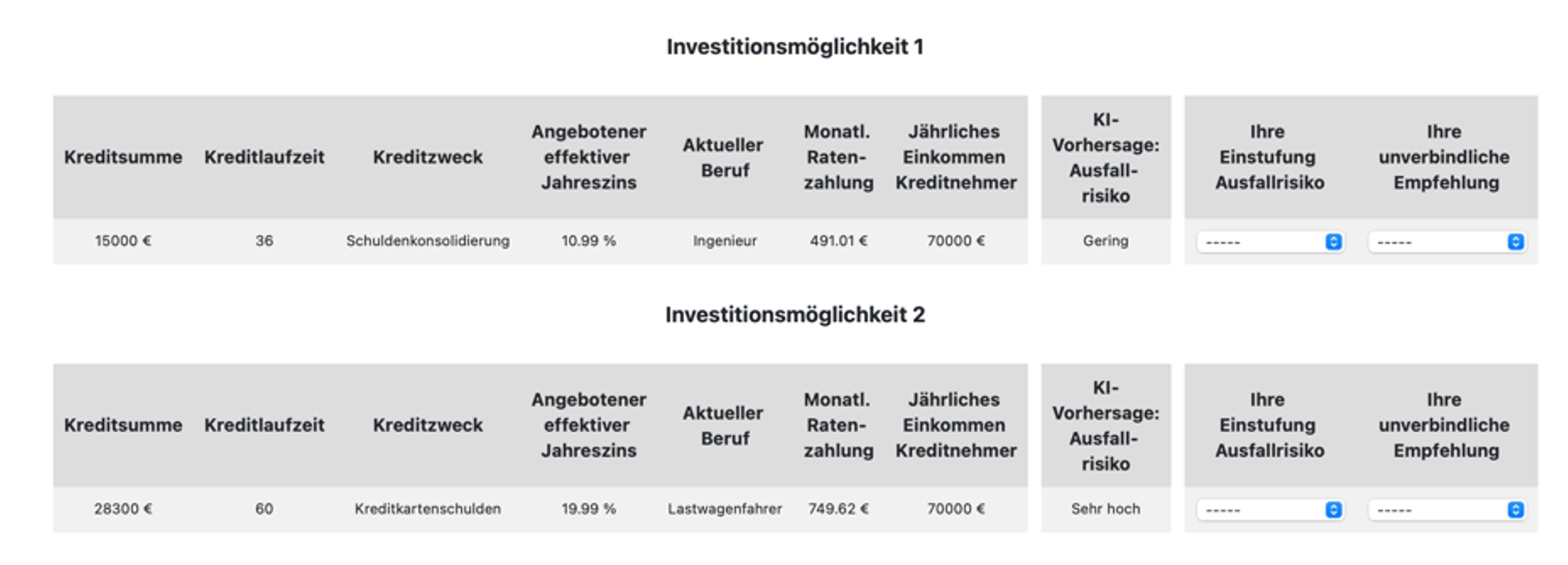}
\end{figure}

\begin{figure}[!h]
    \centering
    \caption{\label{fig:field_interface}The English-translated interface for the ten investment decisions after receiving investment advice}
    \begin{threeparttable}
        \includegraphics[width=\textwidth]{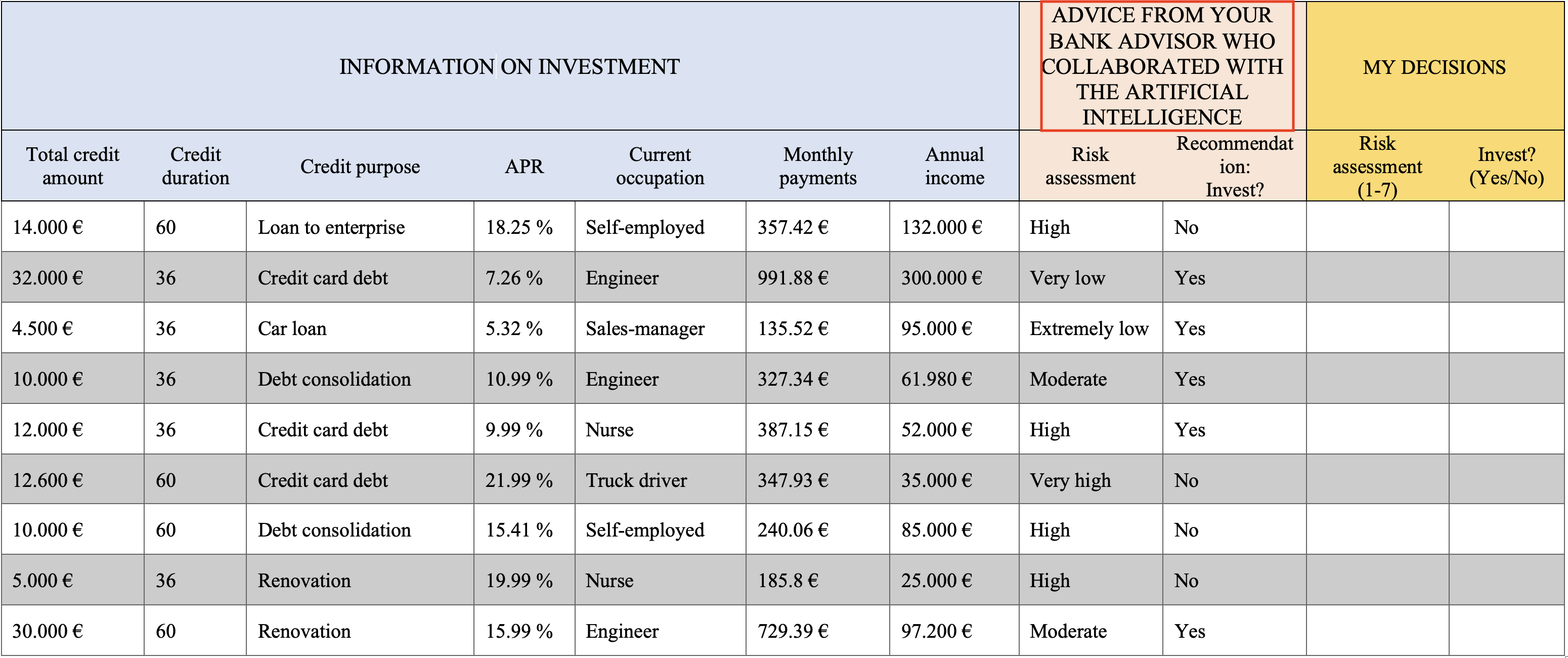}
        \begin{tablenotes}[flushleft]
            \item\leavevmode\kern-\scriptspace\kern-\labelsep\textit{Note}: This specific interface suggests the advice is from a human-AI collaboration. The text in the red box would be ``Your AI advisor" and ``Your savings bank advisor" under the AI-only and Human-only conditions, respectively.            
        \end{tablenotes}
    \end{threeparttable}
    
\end{figure}

\begin{figure}
    \centering
    \caption{The underlying mechanism conceptual model \label{fig:lab_conceptual_model}}
    \includegraphics[width=0.75\linewidth]{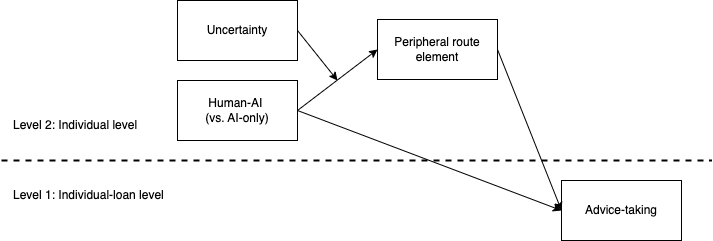}
\end{figure}

\begin{table}[htbp]
        \centering
        \caption{\label{tab:stage1_ai_impact} Bankers' final investment advice is influenced by AI's input.}
        \begin{threeparttable}[b] 
        \footnotesize
        {
\def\sym#1{\ifmmode^{#1}\else\(^{#1}\)\fi}
\begin{tabular}{l*{2}{c}}
\hline\hline
                                                   &DV: GapRiskAssess    &DV: InvestAlign    \\
                                                   &\multicolumn{1}{c}{(1)}    &\multicolumn{1}{c}{(2)}    \\
\hline
AfterAdvice                                        &    -0.485****&     0.119*** \\
                                                   &   (0.081)    &   (0.040)    \\
Constant                                           &     1.419****&     0.585****\\
                                                   &   (0.041)    &   (0.020)    \\
\hline
Observations                                       &       540    &       540    \\
Ind fixed-effects                                  &       Yes    &       Yes    \\
Loan fixed-effects                                 &       Yes    &       Yes    \\
Order fixed-effects                                &       Yes    &       Yes    \\
\hline\hline
\end{tabular}
}

        \begin{tablenotes}[flushleft]
            \scriptsize
            \item\leavevmode\kern-\scriptspace\kern-\labelsep\textit{Note}:
            Standard errors clustered at the banker level are shown in the parentheses; **$p<0.05$, ***$p<0.01$, ****$p<0.001$.
        \end{tablenotes}
        \end{threeparttable}
\end{table}

\begin{table}[htbp]
        \centering
        \caption{\label{tab:loan_advice_details} Ten loan information presented to the bank customers, their default information, and investment advice in the field experiment.}
        \begin{subtable}{\textwidth}
            \centering
            % \scriptsize
            \resizebox{.90\linewidth}{!}{
% Table created by stargazer v.5.2.2 by Marek Hlavac, Harvard University. E-mail: hlavac at fas.harvard.edu
% Date and time: Wed, Jul 17, 2024 - 10:50:07
\begin{tabular}{@{\extracolsep{5pt}} cccccc} 
\\[-1.8ex]\hline 
\hline \\[-1.8ex] 
InvestOrder & 1 & 2 & 3 & 4 & 5 \\ 
\hline \\[-1.8ex] 
LoanAmt (Euro) & 14000 & 32000 &  4500 & 10000 & 12000 \\ 
Term (Months) & 60 & 36 & 36 & 36 & 36 \\ 
InterstRate & 18.25 &  7.26 &  5.32 & 10.99 &  9.99 \\ 
Installment (Euro) & 357.42 & 991.88 & 135.52 & 327.34 & 387.15 \\ 
Income (Euro) & 132000 & 300000 &  95000 &  61980 &  52000 \\ 
BorrowingPurpose & Unternehmensdarlehen & Kreditkartenschulden & Autokredit & Schuldenkonsolidierung & Kreditkartenschulden \\ 
Job & Selbstständig & Ingenieur & Vertrieb & Ingenieur & Pflege \\ 
\hline \\[-1.8ex] 
Actual Default & 0 & 0 & 0 & 0 & 1 \\ 
\hline \\[-1.8ex] 
AI RiskAssess (Stage 1 and 2) & 6 & 2 & 1 & 3 & 3 \\ 
AI InvestRecommend (Stage 2) & 0 & 1 & 1 & 1 & 1 \\ 
Human-AI RiskAssess (Stage 2) & 5 & 2 & 1 & 4 & 5 \\ 
Human-AI InvestRecommend (Stage 2) & 0 & 1 & 1 & 1 & 1 \\ 
\hline 
\end{tabular} 
}
        \end{subtable}
        
        \begin{subtable}{\textwidth}
            \centering
            \resizebox{.9\linewidth}{!}{
% Table created by stargazer v.5.2.2 by Marek Hlavac, Harvard University. E-mail: hlavac at fas.harvard.edu
% Date and time: Wed, Jul 17, 2024 - 10:50:07
\begin{tabular}{@{\extracolsep{5pt}} cccccc} 
\\[-1.8ex]\hline 
\hline \\[-1.8ex] 
InvestOrder & 6 & 7 & 8 & 9 & 10 \\ 
\hline \\[-1.8ex] 
LoanAmt (Euro)& 12600 & 10000 &  5000 & 30000 &  7000 \\ 
Term (Months) & 60 & 60 & 36 & 60 & 36 \\ 
InterstRate & 21.99 & 15.41 & 19.99 & 15.99 &  7.69 \\ 
Installment (Euro)& 347.93 & 240.06 & 185.80 & 729.39 & 218.36 \\ 
Income (Euro)&  35000 &  85000 &  25000 &  97200 &  39000 \\ 
Income (Euro)&  35000 &  85000 &  25000 &  97200 &  39000 \\ 
BorrowingPurpose & Kreditkartenschulden & Schuldenkonsolidierung & Renovierungen & Renovierungen & Kreditkartenschulden \\ 
Job & Lastwagenfahrer & Selbstständig & Pflege & Ingenieur & Team Leader \\ 
\hline \\[-1.8ex] 
ActualDefault & 1 & 1 & 0 & 0 & 0 \\ 
\hline \\[-1.8ex] 
AI RiskAssess (Stage 1 and 2) & 6 & 5 & 4 & 5 & 2 \\ 
AI InvestRecommend (Stage 2) & 0 & 0 & 1 & 0 & 1 \\ 
Human-AI RiskAssess (Stage 2) & 6 & 5 & 5 & 4 & 5 \\ 
Human-AI InvestRecommend (Stage 2) & 0 & 0 & 0 & 1 & 1 \\ 
\hline
\end{tabular} 

}
        \end{subtable}
        
\end{table}

\begin{table}[htpb]
    \centering
    \footnotesize	
    \caption{The description of the advice source for each experimental condition in the field experiment}
    \label{tab:field_source_stimuli}
    {
\def\sym#1{\ifmmode^{#1}\else\(^{#1}\)\fi}
        \begin{tabular}{p{0.2\linewidth}p{0.8\linewidth}}
    \hline\hline
    Advice source & The description of the advice source in English \\
    \hline
    
    AI-only & You will receive a risk assessment and a non-binding investment recommendation from an AI.
    The AI was developed specifically to help you with your decisions. More specifically, based on more than 800,000 individual data points, the AI was developed to accurately predict the borrower's repayment behavior based on the 7 pieces of information displayed. 
    
    The AI we deployed is a so-called Deep Neural Network - one of the most advanced and successful applications of Artificial Intelligence that exist today. Deep Neural Networks are the basis of various everyday applications such as facial recognition on your smartphone, text translation, voice controls, autonomous driving, and much more.
    
    Our tests show that developed AI is very good at correctly identifying those individuals who will not repay an awarded loan. For example, the AI identified 70\% of loan applicants who ultimately defaulted on a loan in a test set. Thus, Artificial Intelligence is designed to help you make a better risk assessment and investment decision. \\ \hline
    
    Human-AI & You receive a risk assessment and a non-binding investment recommendation from a bank manager at your bank who has collaborated with an AI.
    
    The AI was developed specifically to help the bank manager to give you the best possible advice. More specifically, based on more than 800,000 individual data points, the AI was developed to accurately predict the borrower's repayment behavior based on the 7 pieces of information displayed.
    
    The bank manager can combine his/her own expertise with the AI information to provide a more accurate risk assessment and investment recommendation. In this process, it is always the bank manager who decides on the final recommendation to you, never the artificial intelligence. This means that the bank manager is not bound by the assessments of the Artificial Intelligence.
    
    The AI we deployed is a so-called Deep Neural Network - one of the most advanced and successful applications of Artificial Intelligence that exist today. Deep Neural Networks are the basis of various everyday applications such as facial recognition on your smartphone, text translation, voice controls, autonomous driving, and much more.
    
    Our tests show that developed AI is very good at correctly identifying those individuals who will not repay an awarded loan. For example, the AI identified 70\% of loan applicants who ultimately defaulted on a loan in a test set. Thus, AI can help the bank manager make a better risk assessment and investment recommendation.
    
    The bank manager's commission structure is designed to ensure that you only receive assessments and recommendations that are financially in your personal interest. That is, the bank manager has no incentive to mislead you, but wants to help you make a better risk assessment and investment decision.\\ \hline

    Human-only & You will receive a risk assessment and a non-binding investment recommendation from a bank manager from your bank. 
    
    The bank manager made the assessments about the risk and the non-binding recommendation based on the 7 pieces of information about a borrower and his/her experience. The bank manager's commission structure is designed to ensure that you only receive assessments and recommendations that are financially in your personal best interest. That is, the bank manager has no incentive to mislead you, but wants to help you make a better risk assessment and investment decision. 
    \\\hline\hline    
    \end{tabular}
    
}
\end{table}

\begin{table}[htbp]
    \centering
    % \footnotesize
    \caption{Field experiment analyses sample size summary\label{tab:sample}}    
    \resizebox{\textwidth}{!}{{
\def\sym#1{\ifmmode^{#1}\else\(^{#1}\)\fi}
    \begin{tabular}{p{0.26\linewidth} p{0.14\linewidth}p{0.14\linewidth} p{0.2\linewidth} 
                    >{\centering\arraybackslash}p{0.09\linewidth} 
                    >{\centering\arraybackslash}p{0.09\linewidth} 
                    >{\centering\arraybackslash}p{0.09\linewidth} 
                    >{\centering\arraybackslash}p{0.09\linewidth} }
    \hline\hline
        & & & &\multicolumn{4}{c}{Sample Size} \\ 
        \cline{5-8}
        Analysis Type & Includes Loans with Different Advice? & Includes straightliners?& Figure/Table & Human-AI & AI-only & Human-only & Total \\ 
    \hline
        Human-AI (vs. AI-only) & Yes & No & Tables \ref{table:field_main}, \ref{table:field_human}, \ref{table:field_payoff}, \ref{table:field_main_robustness} (Cols 1, 2, 5) & 450 & 419 & -- & 869 \\  
        Human-AI (vs. AI-only) & No & No & Figure \ref{fig:modelfree}, Table \ref{table:field_main_robustness} (Col 4) & 180 & 168 & -- & 348 \\  
        Human-AI (vs. AI-only) & Yes & Yes & Table \ref{table:field_main_robustness} (Col 3) & 460 & 449 & -- & 909 \\  
        Human-only (vs. AI-only) & Yes &No  & Table \ref{table:field_human} & -- & 419 & 430 & 849 \\  
    \hline
    \end{tabular}
    
}}
\end{table}

\begin{table}[htbp]
    \centering
    \caption{Variable name and definition}
    \label{tab:vardef}
    \footnotesize
    {
\def\sym#1{\ifmmode^{#1}\else\(^{#1}\)\fi}
\begin{tabular}{p{10em}p{36em}}
        \hline\hline\\[-2ex] 
        Variable Name & Explanation \\\hline
        \multicolumn{2}{p{42em}}{\textit{Dependent variables}}\\
        
        FinalAlign & Alignment of the final and the advised investment decision on a listing. It takes the value of 1 if the final investment decision is identical to the advised investment decision and 0 otherwise. \\
        GapFinalRiskAssess & The absolute difference between the final and the advised risk assessment on a listing.  \\
        FinalPayoff & The potential payoff if the final investment decision on a listing is selected to reward the customer. The payoff is calculated as the realized monetary payoff contingent on one's final investment decision, following the calculation of $100\times(1-\text{Default})\times \text{Invest} \times (1+\text{APR})^{(\text{Term}/12)} +100\times(1-\text{FinalInvest})$.
        \\ \\
        \multicolumn{2}{p{42em}}{\textit{Customer-loan level variables}}\\
                        
        InitInvest & Initial investment decision on a listing. It takes the value of 1 if invested and 0 otherwise. 
        \\
                        
        InitRiskAssess & Initial risk assessment on a listing. It takes the value between 1 to 7, where 1 and 7 represent the lowest and the highest perceived risk of a customer on a listing, respectively.\\
                                                
        InitAlign & Alignment of the initial and the advised investment decision on a listing. It takes the value of 1 if the initial investment decision is identical to the advised investment decision and 0 otherwise. \\
                                                
        GapInitRiskAssess &  The absolute difference in the $\text{InitRiskAssess}$ and $\text{AdviceRiskAssess}$.   \\
        
        InitPayoff & The potential payoff if the initial investment decision on a listing is selected to reward the customer. The payoff is calculated using the same formula as the one used for the $\text{FinalPayoff}$.\\ \\

        \multicolumn{2}{p{42em}}{\textit{Customer level variables}}\\
                
        Age & Individual customer's age. \\
        
        RiskPref & The stated risk preference of an individual ranges from 0 to 10. \\

        DecisionGrayzoneSize & The range size of risk assessments overlaps for both invested and uninvested loans. For instance, if the maximum risk assessment to invest in a loan is five, and the minimum risk assessment not to invest in a loan is four, with risk assessments of 4 and 5 both being possible for investment and non-investment decisions, then $\text{DecisionGrayzoneSize}$ would have a value of 2. \\

        MinInitRisk & The minimum initial risk assessment for the ten loans. \\

        RangeInitRisk & The absolute difference between the maximum and the minimum initial risk assessment for the ten loans. \\
        Date & The day since the start of the experiment date. \\\\

        \multicolumn{2}{p{42em}}{\textit{Loan-treatment level variables}}\\
        AdviceInvest & The advice on whether to invest in a listing from an advisor. It takes the value of 1 if the advice is to invest and 0 otherwise.\\
                        
        AdviceRiskAssess & The risk assessment about a listing from an advisor, ranging from 1 to 7. \\

        \hline 
\end{tabular}
}

\end{table}

\begin{table}[htbp]
    \centering
    \caption{\label{table:sumstat}Summary statistics of the field experiment variables}
    \footnotesize
    {
\def\sym#1{\ifmmode^{#1}\else\(^{#1}\)\fi}
\begin{tabular}{l*{1}{cccccc}}
\hline\hline
                % &\multicolumn{6}{c}{}                              
                &     Mean&  Std Dev&      Min&   Median&      Max&Observations\\
\hline
FinalAlign      &    0.673&    0.469&        0&        1&        1&     1299\\
GapFinalRiskAssess&    1.144&    1.138&        0&        1&        6&     1299\\
FinalPayoff     &  110.818&   47.608&        0&      100&    231.2&     1299\\
InitInvest      &    0.493&    0.500&        0&        0&        1&     1299\\
InitAlign       &    0.524&    0.500&        0&        1&        1&     1299\\
InitRiskAssess  &    4.066&    1.738&        1&        4&        7&     1299\\
GapInitRiskAssess&    1.564&    1.250&        0&        1&        6&     1299\\
InitPayoff      &  115.152&   54.059&        0&      100&    231.2&     1299\\
Age             &   45.200&   19.518&       16&     47.5&       89&      130\\
RiskPref        &    4.646&    2.206&        0&        5&       10&      130\\
MinInitRisk     &    1.777&    0.883&        1&        2&        6&      130\\
RangeInitRisk   &    4.554&    1.208&        1&        5&        6&      130\\
DecisionGrayzoneSize&    1.485&    1.873&        0&        1&        7&      130\\

Date         &  23.785 &   16.145   &      1 &       23 &       54     &  130 \\

\hline\hline
\end{tabular}
}

\end{table}

\begin{table}[htbp] 
    \centering
        \small
        \caption{Field experiment randomization check \label{tab:field_sum_condition}}
        
            \begin{subtable}{\textwidth}
                \centering
                \footnotesize
                \begin{threeparttable}
                    \caption{Customer level control variables' means (and standard deviation)\label{tab:balance_check_ind}}
                    {
\def\sym#1{\ifmmode^{#1}\else\(^{#1}\)\fi}
\begin{tabular}{l*{2}{c}}
\hline\hline
                    &\multicolumn{1}{c}{(1)}&\multicolumn{1}{c}{(2)}\\
                    &\multicolumn{1}{c}{AI-only}&\multicolumn{1}{c}{Human-AI}\\
\hline
Age                 &      $44.444_a$         &      $45.167_a$                \\
                    &    (19.202)             &    (19.245)                \\
[1em]
Riskpref            &       $4.911_a$         &       $4.786_a$                \\
                    &     (1.987)             &     (2.181)                \\
[1em]
MinInitRisk         &       $1.844_a$         &       $1.881_a$                \\
                    &     (0.767)             &     (0.968)                \\
[1em]
RangeInitRisk   &       $4.444_{a}$         &       $4.333_a$                 \\
                    &     (1.078)            &     (1.373)                 \\
[1em]
DecisionGrayzoneSize&       $1.422_a$         &       $1.667_a$                \\
                    &     (1.983)             &     (1.883)                \\
\hline
Observations        &          45             &          42                \\
\hline\hline
\end{tabular}
}

                \end{threeparttable}
            \end{subtable}

            \vspace{1em}
            
            \begin{subtable}{\textwidth}
                \centering
                \footnotesize
                \begin{threeparttable}
                    \caption{Customer-loan level control variables' means (and standard deviation)\label{tab:balance_check_ind_loan}}
                    {
\def\sym#1{\ifmmode^{#1}\else\(^{#1}\)\fi}

\begin{tabularx}{\textwidth}{p{3cm} *{4}{>{\centering\arraybackslash}X}} 
\toprule
 & \multicolumn{2}{c}{Data based on all loans} & \multicolumn{2}{c}{Data based on loans with identical advice} \\
\cmidrule(lr){2-3} \cmidrule(lr){4-5} 
                    & (1) & (2) & (3) & (4)   \\
                    & AI-only & Human-AI & AI-only & Human-AI \\
\midrule
InitInvest          & $0.516_{a}$  & $0.508_{a}$  & $0.522_{\dagger}$  & $0.512_{\dagger}$  \\
                    & (0.500)      & (0.501)      & (0.501)            & (0.501)            \\
[1em]
InitAlign           & $0.511_{a}$  & $0.537_{a}$  & $0.689_{\dagger}$  & $0.655_{\dagger}$  \\
                    & (0.500)      & (0.499)      & (0.464)            & (0.477)            \\
[1em]
InitRiskAssess      & $4.071_{a}$  & $4.033_{a}$  & $4.133_{\dagger}$  & $4.000_{\dagger}$  \\
                    & (1.686)      & (1.670)      & (1.801)            & (1.831)            \\
[1em]
GapInitRiskAssess   & $1.811_{b}$  & $1.442_{a}$  & $1.722_{\dagger}$  & $1.607_{\dagger}$  \\
                    & (1.365)      & (1.217)      & (1.446)            & (1.414)            \\
[1em]
$\ln$(InitPayoff+1) & $4.410_{a}$  & $4.296_{a}$  & $3.909_{\dagger}$  & $3.850_{\dagger}$  \\
                    & (1.367)      & (1.515)      & (1.755)            & (1.802)            \\
\midrule
Observations        & 450          & 419          & 180                & 168                \\
\bottomrule
\end{tabularx}
}

                \begin{tablenotes}[flushleft]
                    \small
                    \item \textit{Note}: Means (and standard deviation) with different subscripts differ at $p<0.05$ using pair-wise two-sample t-tests.  Please note the absolute difference in initial risk assessments and subsequently observed advice ($GapInitRiskAssess$) was significantly higher in the AI-only condition compared to the Human-AI and Human-only conditions. Since we do not observe any differences in the initial risk assessments ($InitRiskAssess$) between the Human-AI and AI-only conditions, the discrepancy in $GapInitRiskAssess$ is arguably of a technical nature, as the AI-only condition provided different advice for six of the ten investment opportunities. This is further supported by the non-statistically significant difference in $GapInitRiskAssess$ when focusing on the four loans with identical advised risk assessments across conditions ($p=0.45$, two-sample Student's t-test), as shown in Table \ref{tab:balance_check_ind_loan} Columns (3)-(4), which confirms no systematic imbalances before exposure to advice.
                \end{tablenotes}
            \end{threeparttable}
            \end{subtable}

\end{table}

\clearpage

\begin{table}
        \centering
            \caption{\label{table:field_main_robustness}Bank customers are more likely to align their final investment decision with advice under Human-AI than AI-only condition -- field experiment robustness check results}
            \begin{threeparttable}
                \footnotesize
                {
\def\sym#1{\ifmmode^{#1}\else\(^{#1}\)\fi}
\begin{tabular}{l*{5}{c}}
\hline\hline
        &\multicolumn{4}{c}{DV: FinalAlign}&\multicolumn{1}{c}{DV: GapFinalRiskAssess}\\
                                 \cline{2-6}
                                 &   Logistic    & OLS & OLS & OLS &OLS \\
\cline{2-6}
                               &\multicolumn{1}{c}{(1)} &\multicolumn{1}{c}{(2)} &\multicolumn{1}{c}{(3)}  &\multicolumn{1}{c}{(4)}
                               &\multicolumn{1}{c}{(5)}\\
                               
\hline
Human-AI (vs. AI-only)           &               0.984****&               0.154*** &               0.148*** &               0.136*** &              -0.270**  \\
                                 &             (0.283)    &             (0.046)    &             (0.045)    &             (0.042)    &             (0.128)    \\
Straightliner                    &                        &                        &              -0.211****&                        &                        \\
                                 &                        &                        &             (0.053)    &                        &                        \\
Constant                         &              -2.846*** &               0.453****&               0.440****&               0.440****&               1.286****\\
                                 &             (0.950)    &             (0.037)    &             (0.037)    &             (0.059)    &             (0.086)    \\
\hline
Observations                     &                 869    &                 869    &                 909    &                 348    &                 869    \\
\(R^{2}\)                        &                        &               0.244    &               0.268    &               0.301    &               0.312    \\
Pseudo \(R^{2}\)                 &               0.211    &                        &                        &                        &                        \\
Loan $\times$ Advice fixed-effects&                 Yes    &                 Yes    &                 Yes    &                        &                 Yes    \\
Loan fixed-effects               &                        &                        &                        &                 Yes    &                        \\
Date fixed-effects               &                 Yes    &                 Yes    &                 Yes    &                 Yes    &                 Yes    \\
Branch fixed-effects             &                  No    &                 Yes    &                  No    &                  No    &                  No    \\
Controls                         &                 Yes    &                 Yes    &                 Yes    &                 Yes    &                 Yes    \\

\hline\hline
\end{tabular}
}

                \begin{tablenotes}[flushleft]
                    \footnotesize
                    \item\leavevmode\kern-\scriptspace\kern-\labelsep\textit{Notes}: Standard errors clustered at the investor level are shown in the parentheses; **$p<0.05$, ***$p<0.01$, ****$p<0.001$.
            \end{tablenotes}
        \end{threeparttable} 
\end{table}

\begin{table}[htpb]
    \centering
    \footnotesize	
    \caption{The description of the advice source for each experimental condition in the laboratory experiment.}
    \label{tab:lab_source_stimuli}
    {
\def\sym#1{\ifmmode^{#1}\else\(^{#1}\)\fi}
        \begin{tabular}{p{0.2\linewidth}p{0.8\linewidth}}
    \hline\hline
    Advice source & The description of the advice source in English \\
    \hline
    
    AI-only & Please be aware:
    The recommendations you will receive below come from an artificial intelligence developed by the research team, which assesses the risk of investment opportunities in order to make recommendations. This system uses advanced algorithms and data analysis to evaluate potential risks. The final assessments and recommendations therefore always come exclusively from one machine.

    To calculate your earnings from this study, one of your decisions from level 1 or level 2 will be used. Accordingly, the next decisions are also potentially relevant for payment.

    Please consider all provided information carefully before making your decisions, as it may significantly impact your potential earnings from this study. It is essential to weigh all factors and understand the implications thoroughly to maximize your benefits and avoid potential drawbacks.
    \\ \hline
    
    Human-AI & Please be aware:
    The recommendations that you will receive below come from a real, previously interviewed bank advisor to whom we have previously shown the investment opportunities. The bank consultant had access to the artificial intelligence developed by the research team, which assesses the risk of investment opportunities in order to make a recommendation. Nevertheless, the final assessments and recommendations always come from the bank advisor.

    One of your decisions from Level 1 or Level 2 will be used to calculate your earnings from this study. Accordingly, the next decisions are also potentially relevant for payment.

    Please consider all provided information carefully before making your decisions, as it may significantly impact your potential earnings from this study. It is essential to weigh all factors and understand the implications thoroughly to maximize your benefits and avoid potential drawbacks.\\
    \hline\hline    
    \end{tabular}
    
}
\end{table}    
    
\begin{table}[htpb]
        \centering
        \caption{\label{table:perceptual_measure}Items used to measure the perceptual constructs in the laboratory experiment}
        \footnotesize
        {
\def\sym#1{\ifmmode^{#1}\else\(^{#1}\)\fi}
        \begin{tabular}{p{0.2\linewidth}p{0.8\linewidth}}
    \hline\hline
    Construct & Item \\ \hline
    Autonomy & If I follow the recommendation of an artificial intelligence, I feel less likely to give up my decision-making autonomy than if I follow a human's recommendation. \\
    \hline
    Disappointment & Given your initial expectation regarding the accuracy of the recommendations, to what extent were you disappointed with the accuracy of the recommendations (70\%)? \\ \hline
    Responsibility & Who is responsible for your decision? (Options: 1. Me / 2. The advisor / 3. Me and the advisor.) \\ \hline
    % Advisor accountability & 1. The advisor is to be blamed if I make a wrong decision regarding this investment opportunity. \\
    % & 2. The advisor is responsible for me to make a good decision for this investment opportunity.\\
    % \hline
    % Belief in advice quality & 1. I am confident in this investment recommendation for the loan request.\\
    % & 2. I believe the advice is a good investment recommendation for the loan request.\\
    % & 3. The investment recommendation accurately reflects the chance of default for this loan request. \\ 
    % \hline
    
    %Tolerance of wrong recommendations & I get really angry with the advisor if I receive a wrong recommendation. \\ \hline
    
    % Social influence & 1. I do not feel comfortable rejecting the investment recommendation from the advisor even if I disagree with it. \\
    % & 2. I feel bad if I ignore the advisor’s recommendation because it is important to be courteous/polite. \\
    % & 3. I feel a social obligation to follow the recommendation of the advisor.
    % \\ \hline
    CogTrust & 1. The human/AI advisor is like a real expert in predicting the loan default. \\
    & 2. The human/AI advisor is with good knowledge in predicting the loan default. \\
    & 3. The human/AI advisor provides an unbiased estimation of the loan default. \\
    \hline
    EmoTrust & 1. I feel secure about relying on the human/AI 
    advisor for making my investment decision. \\
    & 2. I feel comfortable about relying on the human/AI 
    advisor for making my investment decision. \\
    & 3. I feel content about relying on the human/AI 
    advisor for making my investment decision. \\

    \hline\hline    
    \end{tabular}
    
}
\end{table}

\begin{table}[htpb]
        \centering
        \caption{\label{table:sumstat_lab}Summary statistics of the laboratory experiment variables}
        \footnotesize
        {
\def\sym#1{\ifmmode^{#1}\else\(^{#1}\)\fi}
\begin{tabular}{l*{1}{cccccc}}
\hline\hline
                &     Mean&  Std Dev&      Min&   Median&      Max&Observations\\
\hline
FinalAlign      &    0.679&     0.47&        0&        1&        1&      870\\
GapFinalRiskAssess&    0.600&     0.73&        0&        0&        4&      870\\
InitInvest      &    0.478&     0.50&        0&        0&        1&      870\\
% GapInitInvest&    0.515&     0.50&        0&        1&        1&      870\\
InitAlign &    0.485&     0.50&        0&        0&        1&      870\\
InitRiskAssess  &    3.977&     1.39&        1&        4&        7&      870\\
GapInitRiskAssess&    1.175&     1.02&        0&        1&        5&      870\\
InitCertainty   &    3.376&     0.92&        1&        4&        5&      870\\
FinalCertainty  &    3.511&     0.85&        1&        4&        5&      870\\
Age             &    32.52&     9.72&       18&       29&       67&       87\\
Female          &    0.333&     0.47&        0&        0&        1&       87\\
RiskPref        &    3.724&     1.29&        1&        4&        6&       87\\
MinInitRisk     &    1.908&     0.91&        1&        2&        4&       87\\
RangeInitRisk   &    3.954&     1.40&        0&        4&        6&       87\\
DecisionGrayzoneSize&    0.805&     0.89&        0&        1&        5&       87\\
PriorBeliefAI   &       69&     15.2&       33&       70&       95&       87\\
PriorBeliefHumanAI&    76.25&     16.8&        9&       80&      100&       87\\
PriorBeliefHuman&    65.25&     16.0&       20&       70&       95&       87\\
ExperiencedInvestor&    3.310&     1.47&        1&        3&        7&       87\\
CogTrust        &    4.517&     0.87&        2&     4.50&     6.50&       87\\
EmoTrust        &    4.230&     1.05&        1&        4&     6.50&       87\\
Disappointment  &    3.161&     1.71&        1&        3&        6&       87\\
Autonomy        &    3.299&     1.73&        1&        3&        7&       87\\
AdvisorResp     &    0.172&     0.38&        0&        0&        1&       87\\
\hline\hline
\end{tabular}
}

\end{table}

\clearpage

\subsection{Replication of Field Experiment Results Using Online Experiment Data\label{replication}}
\setcounter{table}{0}
\setcounter{figure}{0}
\renewcommand{\thetable}{B\arabic{table}}
\renewcommand{\thefigure}{B\arabic{figure}}

Table \ref{table:lab_main} Column (1) presents the OLS regression results based on the online experiment data with robust standard errors clustered at the individual level. The findings indicate that participants align their final investments with advice more in the Human-AI condition than in the AI-only condition. 

\begin{table}[htbp]
    \centering
    \caption{\label{table:lab_main}Participants are more likely to align their final investment decisions and risk assessments with advice under Human-AI than AI-only condition -- laboratory experiment OLS regression results}
    \begin{threeparttable}
        \small
        {
\def\sym#1{\ifmmode^{#1}\else\(^{#1}\)\fi}
\begin{tabular}{l*{3}{c}}
\hline\hline
                                 &     \multicolumn{2}{c}{DV: FinalAlign}                            &DV: GapFinalRiskAssess    \\
                                 \cline{2-4}
                                 &\multicolumn{1}{c}{(1)}    &\multicolumn{1}{c}{(2)}    &\multicolumn{1}{c}{(3)}    \\
\hline
Human-AI (vs. AI-only)           &               0.071**  &               0.066**  &              -0.141**  \\
                                 &             (0.030)    &             (0.027)    &             (0.064)    \\
InitInvest                       &               0.000    &               0.003    &               0.074    \\
                                 &             (0.036)    &             (0.036)    &             (0.062)    \\
InitAlign                        &               0.336****&               0.335****&               0.004    \\
                                 &             (0.043)    &             (0.043)    &             (0.052)    \\
InitRiskAssess                   &               0.005    &               0.005    &               0.053    \\
                                 &             (0.012)    &             (0.012)    &             (0.030)    \\
GapInitRiskAssess                &              -0.034**  &              -0.035**  &               0.282****\\
                                 &             (0.016)    &             (0.015)    &             (0.031)    \\
Age                              &              -0.000    &              -0.000    &               0.004    \\
                                 &             (0.002)    &             (0.002)    &             (0.002)    \\
Female                           &               0.029    &               0.026    &               0.086    \\
                                 &             (0.030)    &             (0.029)    &             (0.064)    \\
RiskPref                         &               0.004    &               0.004    &              -0.013    \\
                                 &             (0.011)    &             (0.011)    &             (0.020)    \\
DecisionGrayzoneSize             &               0.019    &               0.015    &               0.030    \\
                                 &             (0.015)    &             (0.014)    &             (0.047)    \\
MinInitRisk                      &              -0.042    &              -0.044**  &              -0.011    \\
                                 &             (0.021)    &             (0.021)    &             (0.055)    \\
PriorBeliefAI                    &              -0.000    &                        &               0.000    \\
                                 &             (0.001)    &                        &             (0.003)    \\
PriorBeliefHumanAI               &               0.001    &                        &              -0.005    \\
                                 &             (0.001)    &                        &             (0.003)    \\
Constant                         &               0.478****&               0.480****&               0.628****\\
                                 &             (0.036)    &             (0.034)    &             (0.057)    \\
\hline
Observations                     &                 870    &                 870    &                 870    \\
\(R^{2}\)                        &               0.422    &               0.422    &               0.269    \\
Loan fixed-effects               &                 Yes    &                 Yes    &                 Yes    \\
% RangeInitRisk                    &                 Yes    &                 Yes    &                 Yes    \\
Date fixed-effects               &                 Yes    &                 Yes    &                 Yes    \\
\hline\hline
\end{tabular}
}

        \begin{tablenotes}[flushleft]
        \footnotesize
            \item\leavevmode\kern-\scriptspace\kern-\labelsep\textit{Notes}: Standard errors clustered at the individual level are shown in the parentheses; **$p<0.05$, ***$p<0.01$, ****$p<0.01$.
        \end{tablenotes}
    
    \end{threeparttable}
    
\end{table}

Recall that all participants were informed in this laboratory experiment that the prediction accuracy of their advisor equals 70\% \textit{after} reporting their prior beliefs, replicating the findings in the field experiment after controlling for the prior beliefs in the banker-AI and the AI advisor's prediction accuracy could help to rule out the possibility that the higher prior belief in the banker-AI than the AI is the main driver of our results.  We note that the greater alignment in the final investment to the advice is present despite controlling for participants' prior belief in the banker-AI and AI advisor, whose coefficient estimates are not significantly different from zero. We further exclude participants' prior belief in the banker-AI, and AI advisor from the analyses, and the results are shown in Column (2) of Table \ref{table:lab_main}. The coefficient estimate of $\beta_1$ shown in Column (2) is not significantly different from that shown in Column (1) of Table \ref{table:lab_main} ($p=0.90$ using a two-tailed Z-test), suggesting no significant influence of (superior) prior belief in the banker-AI to explain participants' greater reliance on human-AI collaborative than AI advice. We also perform a robustness check using an alternative dependent variable, $GapFinalRiskAssess$. The results, displayed in Column (3) of Table \ref{table:lab_main}, suggest greater alignment in final risk assessments with advice, consistent with the field experiment results shown in Column (5) of Table \ref{table:field_main_robustness}, with no significant difference in the coefficient estimates of $\beta_1$ ($p=0.36$, two-tailed Z-test).

To further investigate the replicability of our online experiment, we investigate in the online experiment the role of perceived risk in impacting participants' alignment in their final investments with advice under the Human-AI and AI-only conditions. Following the practice in the field experiment, we conduct a split on the sample based on the advised risk assessment, where relatively high risk is defined when the risk assessment level is greater than 4 (moderate risk) out of 7 levels (ranging from extremely low to extremely high risk). The sample split is less balanced due to a higher proportion of high-risk loans than low-risk loans (six versus four). We apply the same specification shown in Table \ref{table:lab_main} Column (1) to the subsamples of more and less risky investments. We report the results in Table \ref{table:lab_risky}. The results suggest that individuals' higher alignment in the final investments with advice under the Human-AI condition is primarily driven by more risky investments. The results are qualitatively similar to that shown in Table \ref{table:field_moderation}. 

\begin{table}[htbp]
    \centering
    \caption{Participants align their final investments with advice to a greater extent under the Human-AI than the AI-only condition when making more risky investments  -- laboratory experiment OLS regression results\label{table:lab_risky}}
    \begin{threeparttable}
        \small
        {
\def\sym#1{\ifmmode^{#1}\else\(^{#1}\)\fi}
\begin{tabular}{l*{2}{c}}
\hline\hline
                                 & \multicolumn{2}{c}{DV: FinalAlign} \\
                                 \cline{2-3}
                                 &\multicolumn{1}{c}{(1)} &\multicolumn{1}{c}{(2)} \\
                                 &\multicolumn{1}{c}{More Risky Investments}&\multicolumn{1}{c}{Less Risky Investments}\\
\hline
Human-AI (vs. AI)                &               0.089**  &               0.037    \\
                                 &             (0.044)    &             (0.046)    \\
Constant                         &               0.379****&               0.608****\\
                                 &             (0.054)    &             (0.058)    \\
\hline
Observations                     &                 522    &                 348    \\
\(R^{2}\)                        &               0.429    &               0.319    \\
Loan fixed-effects               &                 Yes    &                 Yes    \\
Date fixed-effects               &                 Yes    &                 Yes    \\
Controls                         &                 Yes    &                 Yes    \\
\hline\hline
\end{tabular}
}

        \begin{tablenotes}[flushleft]
        \footnotesize
        \item\leavevmode\kern-\scriptspace\kern-\labelsep\textit{Notes}: Standard errors clustered at the individual level are shown in the parentheses; ** $p<0.05$, *** $p<0.01$, **** $p<0.001$.
        
    \end{tablenotes}
    \end{threeparttable}
\end{table}

Since we have a more direct measure of uncertainty on the individual level, we also perform a median split based on the uncertainty measure to categorize participants into groups experiencing greater and lower levels of uncertainty. Using the specification outlined in Table \ref{table:lab_main} Column (1), we analyze these subsamples separately. The results, presented in Table \ref{table:lab_uncertain}, indicate that the alignment of final investment decisions with advice in the Human-AI condition is significantly stronger among participants experiencing greater uncertainty. The results shown in Table \ref{table:lab_uncertain} are also qualitatively similar to those shown in Table \ref{table:lab_risky}. 

\begin{table}[htbp]
    \centering
    \caption{Participants align their final investments with advice to a greater extent under the Human-AI than the AI-only condition when experiencing greater uncertainty -- laboratory experiment OLS regression results\label{table:lab_uncertain}}
    \begin{threeparttable}
        \small
        {
\def\sym#1{\ifmmode^{#1}\else\(^{#1}\)\fi}
\begin{tabular}{l*{2}{c}}
\hline\hline
                                 &    \multicolumn{2}{c}{DV: FinalAlign}   \\ \cline{2-3}
                                 &\multicolumn{1}{c}{(1)} &\multicolumn{1}{c}{(2)} \\
                                 &\multicolumn{1}{c}{More Uncertain Investors}&\multicolumn{1}{c}{Less Uncertain Investors}\\
\hline
Human-AI (vs. AI-only)                &               0.106**  &               0.020    \\
                                 &             (0.043)    &             (0.040)    \\
Constant                         &               0.427****&               0.531****\\
                                 &             (0.050)    &             (0.056)    \\
\hline
Observations                     &                 420    &                 450    \\
\(R^{2}\)                        &               0.430    &               0.447    \\
Loan fixed-effects               &                 Yes    &                 Yes    \\
Date fixed-effects               &                 Yes    &                 Yes    \\
Controls                         &                 Yes    &                 Yes    \\
\hline\hline
\end{tabular}
}

        \begin{tablenotes}[flushleft]
        \footnotesize
        \item\leavevmode\kern-\scriptspace\kern-\labelsep\textit{Notes}: Standard errors clustered at the individual level are shown in the parentheses; ** $p<0.05$, *** $p<0.01$, **** $p<0.001$.
        
    \end{tablenotes}
    \end{threeparttable}
\end{table}

\end{document}